\documentclass[11pt]{tep}
\usepackage{dyfact}

\begin{document}

\thispagestyle{empty}
\def\thefootnote{\fnsymbol{footnote}}
\setcounter{footnote}{1}
\null
\mbox{}\hfill  
FR-PHENO-2015-013, ZU-TH~40/15, TTK-15-36   \\
\vskip 0cm
\vfill
\begin{center}
  {\Large \boldmath{\bf 
      Dominant mixed QCD--electroweak \order{\alphas\alpha} corrections to \\
      Drell--Yan processes in the resonance region
    }
    \par} \vskip 2.5em
  {\large
    {\sc Stefan Dittmaier$^{1}$, Alexander Huss$^{2,3}$ \\[.3em]
      and Christian Schwinn$^{1,4}$
    }\\[1ex]
    {\normalsize \it 
      $^1$ Albert-Ludwigs-Universit\"at Freiburg, Physikalisches Institut, \\
      D-79104 Freiburg, Germany \\
      $^2$ Institute for Theoretical Physics, ETH, CH-8093 Z\"urich, Switzerland \\
      $^3$ Department of Physics, University of Z\"urich, CH-8057
      Z\"urich, Switzerland\\
      $^4$Institute for Theoretical Particle Physics and Cosmology,\\
      RWTH Aachen University, D-52056 Aachen
    }
    \\[2ex]
  }
  \par \vskip 1em
\end{center}\par
\vskip .0cm \vfill {\bf Abstract:} \par
A precise theoretical description of $\PW$- and $\PZ$-boson production in
the resonance region is essential for the correct interpretation of
high-precision measurements of the W-boson mass and the effective weak
mixing angle. Currently, the largest unknown fixed-order contribution
is given by the mixed QCD--electroweak corrections of
$\order{\alphas\alpha}$.
We argue, using the framework of the pole expansion for the NNLO
QCD--electroweak corrections established in a
previous paper, that the numerically dominant corrections arise from
the combination of  large QCD corrections to the
 production with the large  
electroweak corrections to the decay of the $\PW$/$\PZ$
 boson. We calculate these so-called factorizable corrections of
 ``initial--final'' type and estimate the impact on the $\PW$-boson mass extraction.
We compare our results to simpler approximate
 combinations of 
electroweak and QCD corrections in terms of
naive products of NLO QCD and 
electroweak correction factors and using  leading-logarithmic approximations 
for QED final-state radiation
as provided by the structure-function approach or QED parton-shower programs.
We also  compute corrections of ``final--final''
 type, which are given by finite counterterms to the leptonic 
vector-boson decays and are found to be numerically negligible. 
\par
\vskip 1cm
\noindent
November 2015
\par
\null
\setcounter{page}{0}
\clearpage
\def\thefootnote{\arabic{footnote}}
\setcounter{footnote}{0}


\section{Introduction}
\label{sec:intro}

The class of Drell--Yan-like processes 
is one of the most prominent types of particle reactions
at hadron colliders and describes the production of a lepton pair through an intermediate gauge-boson decay,
\begin{equation*}
  \Pp\Pp / \Pp\Pap \;\to\; V \;\to\; \Pl_1\Pal_2 \;+\; X .
\end{equation*}
Depending on the electric charge of the colour-neutral gauge boson $V$, the process can be further classified into the neutral-current ($V=\PZ/\Pgg$) and the charged-current ($V=\PWpm$) processes. 
The large production rate in combination with the clean experimental signature of the leptonic vector-boson decay allows this process to be measured with great precision.
Moreover, the Dell--Yan-like production of $\PW$ or $\PZ$ bosons is one of the theoretically best understood and most precisely predicted processes.
As a consequence, electroweak (EW) 
gauge-boson production is among the most important 
``standard-candle'' processes at the LHC (see, e.g.\ Refs.~\cite{Abdullin:2006aa,Gerber:2007xk}).
Its cross section can be used as a luminosity monitor, and the measurement of
the mass and width of the $\PZ$ boson represents a powerful tool for detector
calibration.
Furthermore, the \PW charge asymmetry and the rapidity distribution of the $\PZ$ boson deliver important constraints in the fit of the 
parton distribution functions (PDFs)~\cite{Boonekamp:2009yd}, which represent crucial ingredients for almost all predictions at the LHC.

Of particular relevance for precision tests of the Standard Model 
is the potential of the Drell--Yan process at the LHC  for high-precision measurements in the resonance regions, where the effective weak mixing angle, quantified by $\sin^2\theta_\text{eff}^{\Pl}$,
might be extracted from data with LEP precision~\cite{Haywood:1999qg}.
The $\PW$-boson mass can be determined from a fit to the distributions of the lepton transverse momentum ($p_{\rT,\Pl}$) and the transverse mass of the lepton pair ($M_{\rT,\Pl\Pgn}$) which exhibit Jacobian peaks around $\MW$ and $\MW/2$, respectively, and allow for a precise extraction of the mass
with a sensitivity below \SI{10}{\MeV}~\cite{Besson:2008zs,Baak:2013fwa} provided that PDF uncertainties can be reduced\cite{Rojo:2013nia,Quackenbush:2015yra,Bozzi:2015hha,Bozzi:2015zja}.

To fully exploit the potential of the extraordinary experimental precision that is achievable for the Drell--Yan process, it is necessary to have theoretical predictions that match or even surpass the 
expected accuracy.
The current state of the art includes
QCD corrections at next-to-next-to-leading-order (NNLO)
accuracy~\cite{Hamberg:1990np,Harlander:2002wh,Anastasiou:2003ds,Melnikov:2006di,Melnikov:2006kv,Catani:2009sm,Gavin:2010az,Gavin:2012sy}
supplemented by leading higher-order soft-gluon
effects~\cite{Moch:2005ba,Laenen:2005uz,Idilbi:2005ky,Ravindran:2007sv}
 and soft-gluon resummation for  small transverse momenta~\cite{Balazs:1997xd,Landry:2002ix,Bozzi:2010xn,Mantry:2010mk,Becher:2011xn,Guzzi:2013aja,Catani:2015vma}.
For event generation, next-to-leading-order (NLO) 
calculations  have been matched to parton showers~\cite{Frixione:2006gn,Alioli:2008gx,Hamilton:2008pd}, with a recent effort to include NNLO corrections in a parton-shower framework~\cite{Hoeche:2014aia,Karlberg:2014qua,Alioli:2015toa}.
Concerning EW effects, the 
NLO corrections~\cite{Baur:1997wa,Zykunov:2001mn,Baur:2001ze,Dittmaier:2001ay,Baur:2004ig,Arbuzov:2005dd,CarloniCalame:2006zq,Zykunov:2005tc,CarloniCalame:2007cd,Arbuzov:2007db,Brensing:2007qm,Dittmaier:2009cr}
as well as leading higher-order effects from multiple photon emission and
universal weak effects~\cite{Baur:1999hm,Placzek:2003zg,CarloniCalame:2003ux,Brensing:2007qm,Dittmaier:2009cr} are known. The sensitivity to the photon PDF through 
photon-induced production channels has been studied in 
Refs.~\cite{Arbuzov:2007kp,CarloniCalame:2007cd,Dittmaier:2009cr,Boughezal:2013cwa}.
 
In addition to the N$^3$LO QCD corrections, the next frontier in theoretical fixed-order computations is given by the calculation of the mixed QCD--EW corrections of \order{\alphas\alpha}~\cite{Andersen:2014efa}.
These corrections can affect observables relevant for the \MW determination at the percent level~\cite{Balossini:2009sa} and therefore must be under theoretical control.
Up to now, QCD and EW corrections have been combined in various
approximations~\cite{Cao:2004yy,Richardson:2010gz,Bernaciak:2012hj,Barze:2012tt,Li:2012wna,Barze:2013yca}.  
However, a full NNLO calculation at \order{\alphas\alpha} is necessary 
for a proper combination of  QCD and EW corrections without ambiguities.
Here some partial results for two-loop amplitudes~\cite{Kotikov:2007vr,Kilgore:2011pa,Bonciani:2011zz} as well as the  full \order{\alphas\alpha} corrections to the \PW and \PZ decay widths~\cite{Czarnecki:1996ei,Kara:2013dua} are known. 
A complete calculation of the \order{\alphas\alpha} corrections requires to combine the double-virtual corrections with 
 the \order{\alpha} EW corrections to $\PW/\PZ+\jet$ production~\cite{Kuhn:2005az,Kuhn:2007cv,Hollik:2007sq,Denner:2009gj,Denner:2011vu,Denner:2012ts,Hollik:2015pja}, the \order{\alphas} QCD corrections to $\PW/\PZ+\Pgg$ 
production~\cite{Smith:1989xz,Ohnemus:1992jn,Ohnemus:1994qp,Dixon:1998py,Campbell:1999ah,DeFlorian:2000sg,Hollik:2007sq,Campbell:2011bn,Denner:2014bna,Denner:2015fca}, and the double-real corrections using a method to regularize 
infrared (IR) singularities.

 In a previous paper~\cite{Dittmaier:2014qza}, we have initiated the
 calculation of the \order{\alphas\alpha} corrections to Drell--Yan
 processes in the resonance region via the so-called \emph{pole
   approximation}~(PA)~\cite{Stuart:1991xk}, which has been
 successfully applied to the EW corrections to 
 $\PW$~production~\cite{Wackeroth:1996hz,Baur:1998kt,Dittmaier:2001ay,Dittmaier:2014qza}
 and $\PZ$~production~\cite{Dittmaier:2014qza} at NLO.
 It is based on a systematic expansion of the cross section about the
 resonance pole and is suitable for theoretical predictions in the
 vicinity of the gauge-boson resonance, where the higher precision is
 especially relevant.  The PA splits the corrections into distinct
 well-defined subsets, which can be calculated separately. This allows
 to assess the numerical impact of different classes of corrections
 and to identify the dominant contributions.  More precisely, the
 contributions can be classified into two types: the factorizable and
 the non-factorizable corrections.  In the former, the
corrections can be separately attributed to the production and
the subsequent decay of the gauge boson, whereas in the latter 
the production and decay subprocesses are
 linked by the exchange of soft photons.  
At \order{\alpha}, the PA shows agreement with the known 
NLO EW corrections up to fractions of $1\%$ near the resonance, i.e.\ at a phenomenologically satisfactory level~\cite{Dittmaier:2014qza}.
In particular, the bulk of the NLO EW corrections near the resonance can be attributed to the factorizable corrections to the \PW/\PZ~decay subprocesses, while
the factorizable corrections to the production process are mostly suppressed below the 
percent level,
and the non-factorizable contributions being even smaller.

Based on the quality of the PA at NLO 
we are confident that this approach is suitable to calculate the  $\order{\alphas\alpha}$ corrections with sufficient accuracy for the description of observables that are dominated by the resonances.
 The non-factorizable
 corrections comprise the conceptually most challenging contribution
 to the PA and have been computed at \order{\alphas\alpha} in
 Ref.~\cite{Dittmaier:2014qza}.  They turn out to be very small and,
 thus, demonstrate that for phenomenological purposes the
 \order{\alphas\alpha} corrections can be factorized into terms
 associated with initial-state and/or final-state corrections and combinations of the two types.
 In this paper we calculate the factorizable corrections of the type
 ``initial--final'', which combine large QCD corrections to the
 production with the large EW corrections to the decay of the $\PW$/$\PZ$
 boson. 
 Therefore we expect to capture  the dominant contribution at
 \order{\alphas\alpha} to observables relevant for precision physics
 dominated by the \PW and \PZ resonances.  We also compute the corrections of ``final--final'' type, which are given only by finite counterterms to the leptonic vector-boson decay. 
 The remaining factorizable ``initial--initial''
 corrections are expected to deliver only a small contribution 
and would further require $\order{\alphas\alpha}$-corrected PDFs for a consistent evaluation, which are however not available.
It is all the more important to isolate this contribution in a well-defined
manner, as it is accomplished by the PA.

A technical aspect of higher-order calculations involving massless particles is the proper treatment of 
IR singularities that are associated with configurations involving soft and/or collinear particles.
To this end, we use the dipole subtraction 
formalism~\cite{Catani:1996vz,Dittmaier:1999mb,Catani:2002hc,Dittmaier:2008md} 
and its extension for decay processes presented in Ref.~\cite{Basso:2015gca}
for the analytic cancellation of all IR singularities.
Although the cancellation of IR singularities in the \order{\alphas\alpha} corrections presented in this work is accomplished by using a combined approach of the techniques developed for NLO calculations, it represents one of the main technical difficulties in the calculation and we devote special attention to its discussion.

This paper is organized as follows:
In Section~\ref{sec:nnlo} we present the calculation of the initial--final and
final--final factorizable corrections.
We discuss the construction of an IR-finite final result for the
initial--final corrections in detail with a special focus on the treatment of
the combined IR singularities of the QCD and EW corrections. 
Our numerical results are presented in Section~\ref{sec:IF:numerics}, where we compare them to different versions of a naive product ansatz obtained by multiplying NLO QCD and EW correction 
factors, and to a leading-logarithmic treatment of photon radiation 
as provided by the structure-function approach or QED parton showers such as PHOTOS~\cite{Golonka:2005pn}.
We further perform a $\chi^2$ fit in order to estimate the effect of the NNLO
\order{\alphas\alpha} corrections on the measurement of the  $\PW$-boson mass.
A summary is given in Sect.~\ref{sec:concl}.


\section{Calculation of the dominant \texorpdfstring{$\order{\alphas\alpha}$}{O(as a)} corrections in pole approximation}
\label{sec:nnlo}

In this section we identify and calculate the dominant $\order{\alphas\alpha}$ corrections to the charged-current and neutral-current Drell--Yan processes in the vicinity of an intermediate vector-boson resonance. 
In Sect.~\ref{sec:method} we describe the classification of the  $\order{\alphas\alpha}$ corrections in the framework of the PA~\cite{Dittmaier:2014qza}. We identify  factorizable contributions of ``initial--final'' type---i.e.\ the combination of QCD corrections to vector-boson production with EW corrections to vector-boson decay---as dominant source for corrections to distributions dominated by the vector-boson resonance.
The calculation of the building blocks contributing to 
the initial--final factorizable corrections is performed in Sect.~\ref{sec:calc-fact-nnlo-if}.
In Sect.~\ref{sec:IF:master} the different building blocks of the initial--final contributions are combined into a formula suitable for numerical evaluation, where all IR singularities are cancelled explicitly.
Finally, in Sect.~\ref{sec:calc-fact-nnlo-ff} we calculate corrections of ``final--final'' type, which are given by pure counterterm contributions and are numerically small.

\subsection{Survey of types of \texorpdfstring{$\order{\alphas\alpha}$}{O(as a)} corrections in pole approximation}
\label{sec:method}

\begin{figure}[b]
  \centering
  \begin{subfigure}[m]{.48\linewidth}
    \centering
    \includegraphics{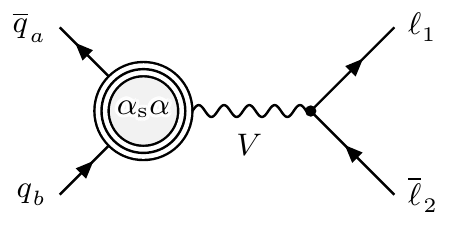}
    \subcaption{Factorizable initial--initial corrections}
  \end{subfigure}
  \begin{subfigure}[m]{.48\linewidth}
    \centering
    \includegraphics{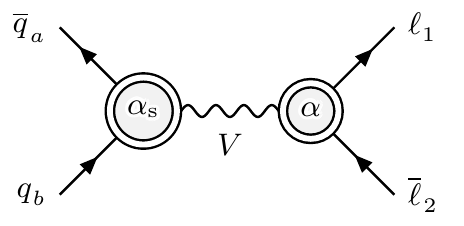}
    \subcaption{Factorizable initial--final corrections}
  \end{subfigure}
  \\[1.5em]
  \begin{subfigure}[m]{.48\linewidth}
    \centering
    \includegraphics{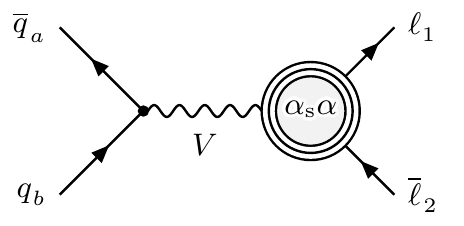}
    \subcaption{Factorizable final--final corrections}
  \end{subfigure}
  \begin{subfigure}[m]{.48\linewidth}
    \centering
    \includegraphics{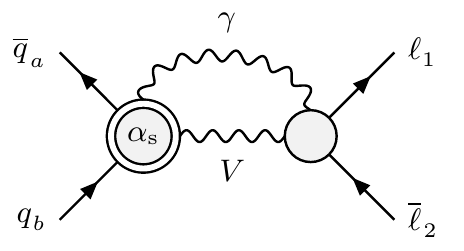}
    \subcaption {Non-factorizable corrections}
  \end{subfigure}
  \caption{The four types of corrections that contribute to the mixed QCD--EW corrections in the PA illustrated in terms of generic two-loop amplitudes. Simple circles symbolize tree structures, double circles one-loop corrections, and triple circles two-loop contributions.}
  \label{fig:NNLOcontrib}
\end{figure}
The PA for Drell--Yan
processes~\cite{Stuart:1991xk,Wackeroth:1996hz,Baur:1998kt,Dittmaier:2001ay,Dittmaier:2014qza}
provides a systematic classification of contributions to Feynman
diagrams that are enhanced by the resonant propagator of a vector
boson~$V=\PW,\PZ$.
The leading corrections
in the expansion around the resonance pole arise from factorizable
corrections to $\PW$/$\PZ$ production and decay subprocesses, and
non-factorizable corrections that link production and decay by
soft-photon exchange.  The PA separates corrections to production and
decay stages in a consistent and gauge-invariant way.  
This is
particularly relevant for the charged-current Drell-Yan process, where
photon radiation off the intermediate $\PW$~boson contributes
simultaneously to the corrections to production and decay of a
$\PW$~boson, and to the non-factorizable contributions.  Applications
of different variants of the PA to NLO EW
corrections~\cite{Wackeroth:1996hz,Baur:1998kt,Dittmaier:2001ay,Dittmaier:2014qza}
have been validated by a comparison to the complete EW NLO
calculations and show excellent agreement at the order of some $0.1\%$
in kinematic distributions dominated by the resonance region.

The structure of the PA for 
the \order{\alphas\alpha} correction has
been worked out in Ref.~\cite{Dittmaier:2014qza}, where details of the
method and our setup can be found.  
The corrections can be classified into the four types of
contributions shown in Fig.~\ref{fig:NNLOcontrib} for the case of the
double-virtual corrections. For each class of contributions with
the exception of the final--final corrections (c), also the
associated real--virtual and 
double-real corrections have to be
computed, obtained by replacing one or both of the labels $\alpha$ and $\alphas$ in
the blobs in Fig.~\ref{fig:NNLOcontrib} by a real
photon or gluon, respectively. The corresponding crossed partonic channels, e.g.\ with
quark--gluon initial states have to be included in addition.

In detail, the four types of corrections are characterized as follows:
\begin{enumerate}[(a)]
\item The initial--initial factorizable corrections are given by
  two-loop \order{\alphas\alpha} corrections to on-shell $\PW/\PZ$
  production and the corresponding one-loop real--virtual and
  tree-level double-real contributions, i.e.\ $\PW/\PZ+\jet$
  production at \order{\alpha}, $\PW/\PZ+\Pgg$ production at
  \order{\alphas}, and the processes $\PW/\PZ+\Pgg+\jet$ at tree
  level. Results for individual ingredients of the initial--initial
  part are known, such as partial two-loop
  contributions~\cite{Kotikov:2007vr,Bonciani:2011zz} and the full
  \order{\alpha} EW corrections to W/Z+jet production including the
  W/Z decays~\cite{Denner:2009gj,Denner:2011vu,Denner:2012ts}.
  However, a consistent combination of these building blocks requires
  also a subtraction scheme for IR singularities at ${\cal
    O}(\alphas\alpha)$ and has not been performed yet.  Note that
  currently no PDF set including \order{\alphas\alpha} corrections is
  available, which is required to absorb IR singularities of the
  initial--initial corrections from photon radiation collinear to the
  beams.

  Results of the PA at \order{\alpha} show that observables such as
  the transverse-mass distribution in the case of \PW production or
  the lepton-invariant-mass distributions for \PZ production are
  extremely insensitive to initial-state photon
  radiation~\cite{Dittmaier:2014qza}. Since these distributions also
  receive relatively moderate QCD corrections, we do not expect
  significant initial--initial NNLO
  \order{\alphas\alpha}
  corrections to such distributions. For observables sensitive to
  initial-state recoil effects, such as the transverse-lepton-momentum
  distribution, 
the \order{\alphas\alpha} corrections should be larger, but still
very small compared to the huge QCD 
corrections.%
\footnote{Note that for such observables
  a fixed-order QCD description is not adequate near the
  Jacobian peak, so that in this case the initial--initial corrections need
  to be combined with a resummation of multiple gluon emissions.  At
  present, such a resummation is available in the POWHEG framework in
  combination with an approximation to the double-real and
  real--virtual part of the initial--initial corrections where the
  first emitted photon or gluon is treated exactly, while further
  emissions are generated in the collinear
  approximation~\cite{Barze:2012tt,Barze:2013yca}.}

\item  The factorizable initial--final corrections 
consist of the \order{\alphas} corrections to $\PW/\PZ$ production combined with the \order{\alpha} corrections to the leptonic  $\PW/\PZ$ decay.
Both types of corrections are large and have a sizable impact on the shape of differential distributions at  NLO, so that we
 expect this class of the factorizable corrections to capture the dominant  \order{\alphas\alpha} effects.
The computation of these contributions is the main result of this paper and is discussed in Sect.~\ref{sec:calc-fact-nnlo-if}.  
 Preliminary numerical results of these corrections were presented in
 Refs.~\cite{Dittmaier:2014koa,Huss:2014eea}.

\item Factorizable final--final corrections arise from the
  $\order{\alphas\alpha}$ counterterms of the
  lepton--$\PW/\PZ$-vertices, which involve only QCD corrections to
  the vector-boson self-energies. There are no corresponding real
  contributions, so that
the final--final corrections have practically no
  impact on the shape of distributions. We compute these corrections
  in Sect.~\ref{sec:calc-fact-nnlo-ff} below and confirm the
  expectation that they are phenomenologically negligible.

\item  The non-factorizable $\order{\alphas\alpha}$ corrections are given by soft-photon corrections
connecting the initial state, the intermediate vector boson, and the
final-state leptons, combined with QCD corrections to $V$-boson production. 
As shown in detail in Ref.~\cite{Dittmaier:2014qza}, these corrections
can be expressed in terms of soft-photon
correction factors to squared tree-level or one-loop QCD matrix
elements by using gauge-invariance arguments.
The numerical impact of these corrections was found to be below the
$0.1\%$ level and is therefore negligible for all phenomenological purposes.
\end{enumerate}

\begin{figure}
  \centering
  \begin{subfigure}[b]{\linewidth}
    \centering
\includegraphics[scale=1]{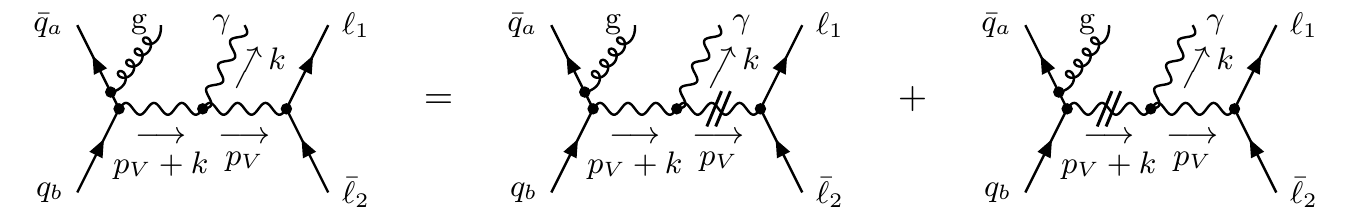} 
\caption{Decomposition of a diagram with photon emission off a $V$-boson line into initial--initial and initial--final corrections}
\label{fig:amp_NNLO:propagator}
      \end{subfigure}
  \begin{subfigure}[b]{.5\linewidth}
  \centering
  \includegraphics[scale=1.]{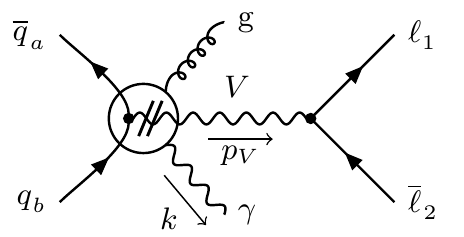}
  \caption{Photon emission from the production subprocess}
  \label{fig:amp_NNLO:production}
  \end{subfigure}
  \hfill
  \begin{subfigure}[b]{.475\linewidth}
  \centering
  \includegraphics[scale=1.]{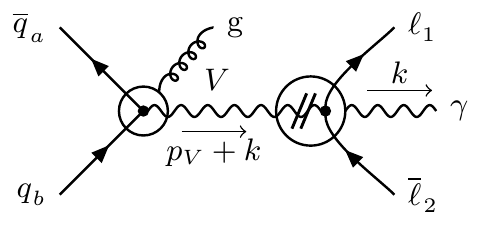}
  \caption{Photon emission from the decay subprocess}
    \label{fig:amp_NNLO:decay}
  \end{subfigure}\hfill
  \caption{
Decomposition of the double-real corrections $\Paq_a(p_a)\Pq_b(p_b)\to\Pl_1(k_1)\Pal_2(k_2)\Pg(k_{\Pg})\Pgg(k)$  
into initial---initial~(\subref{fig:amp_NNLO:production}) and initial--final~(\subref{fig:amp_NNLO:decay}) parts,
illustrated for an example in part~(\subref{fig:amp_NNLO:propagator}). 
  The momentum $p_V$ of the intermediate vector boson $V$ is given by $p_V=p_a+p_b-k_{\Pg}-k=k_1+k_2$.  A double line on a $V$ propagator indicates on-shellness while a gauge boson attached to an  encircled subdiagram indicates all possible insertions.}
 \label{fig:amp_NNLO}
\end{figure}
\begin{figure}
  \centering
  \begin{subfigure}[m]{\linewidth}
    \centering
    \includegraphics[scale=0.95]{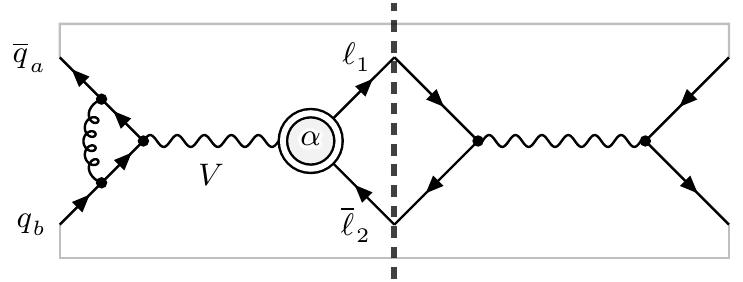}
    \qquad
    \includegraphics[scale=0.95]{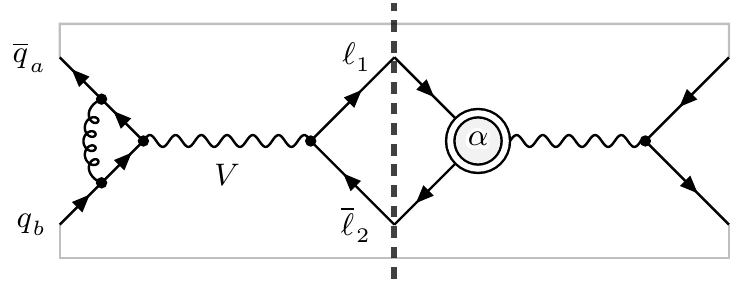}
    \caption{Factorizable initial--final double-virtual corrections}
    \label{fig:NNLO-if-graphs-vv}
  \end{subfigure}
  \\[1.5em]
  \begin{subfigure}[m]{\linewidth}
    \centering
    \includegraphics[scale=0.95]{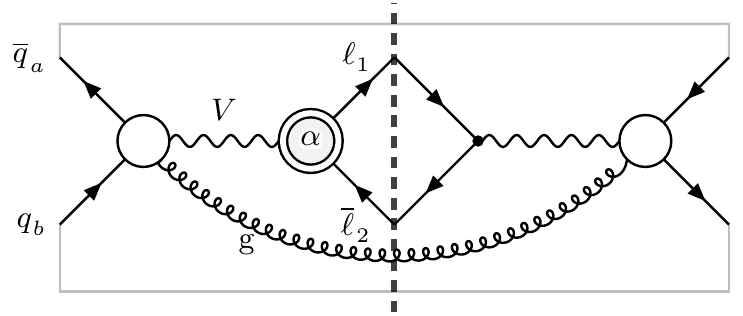}
    \qquad
    \includegraphics[scale=0.95]{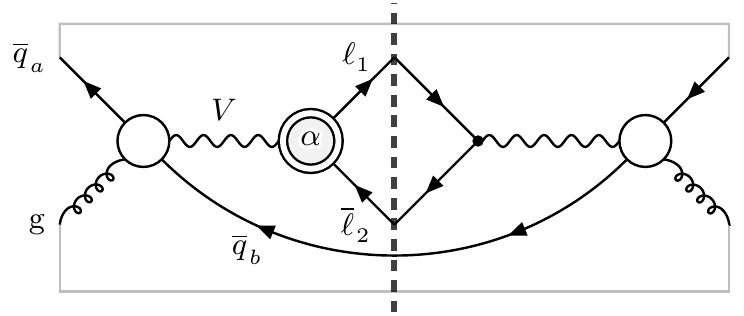}
    \caption{Factorizable initial--final (real QCD)$\times$(virtual EW) corrections}
    \label{fig:NNLO-if-graphs-rv}
  \end{subfigure}
  \\[1.5em]
  \begin{subfigure}[m]{\linewidth}
    \centering
    \includegraphics[scale=0.95]{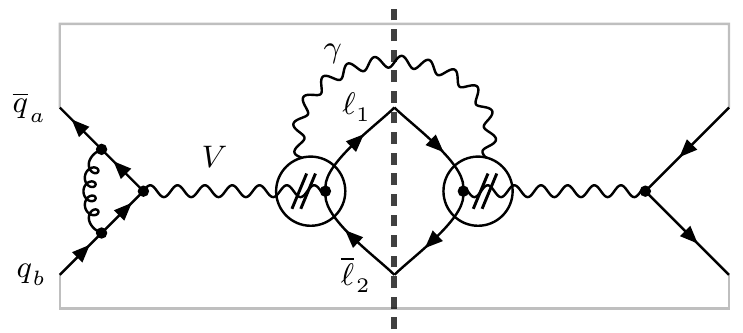}
    \caption{Factorizable initial--final (virtual QCD)$\times$(real photonic) corrections}
    \label{fig:NNLO-if-graphs-vr}
  \end{subfigure}
  \\[1.5em]
  \begin{subfigure}[m]{\linewidth}
    \centering
    \includegraphics[scale=0.95]{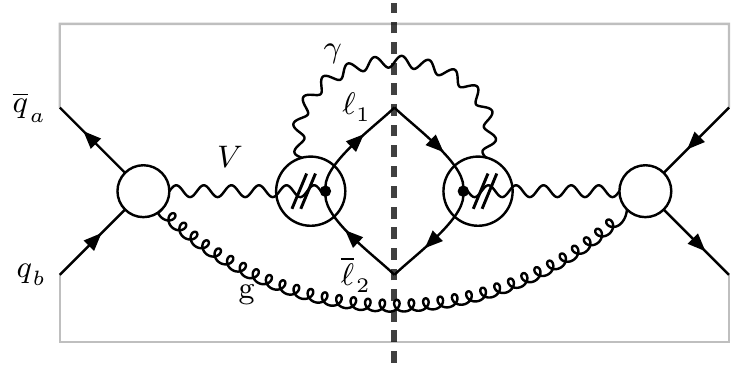}
    \qquad
    \includegraphics[scale=0.95]{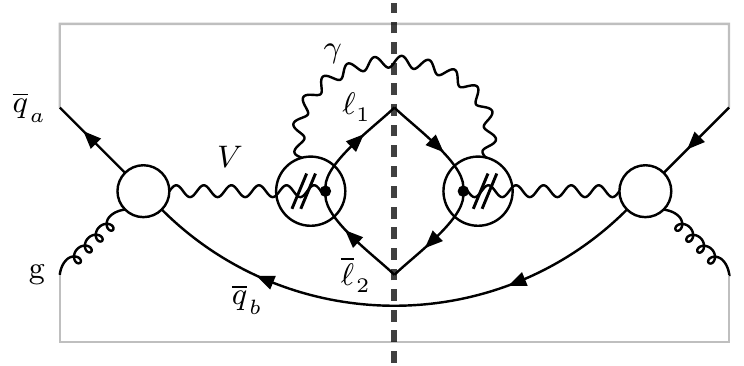}
    \caption{Factorizable initial--final double-real corrections}
    \label{fig:NNLO-if-graphs-rr}
  \end{subfigure}
  \caption{Interference diagrams for the various contributions to the factorizable initial--final corrections of \order{\alphas\alpha}, with blobs representing all relevant tree structures.
  The blobs with ``\alpha'' inside represent one-loop corrections of \order{\alpha}.}
  \label{fig:NNLO-if-graphs}
\end{figure}

The definition of the factorizable corrections and the separation of
initial- and final-state corrections is illustrated in
Fig.~\ref{fig:amp_NNLO} for the case of the double-real
corrections. An example
diagram for the charged-current process is given in Fig.~\ref{fig:amp_NNLO:propagator}, which
cannot be attributed uniquely to the vector-boson production or decay
subprocess and displays an overlapping resonance structure due to the
propagator poles at $p_V^2=\mu_V^2$ and $(p_V+k)^2=\mu_V^2$.
Here $\mu_V$ combines the real mass and width parameters of $V$,
$M_V$ and $\Gamma_V$, to a complex mass value,
$\mu_V^2=M_V^2-\ri M_V\Gamma_V$.
However,
a simple partial-fractioning identity for the two $V$-boson
propagators allows us
to disentangle the two resonance structures and to
decompose such diagrams into contributions associated with photon
emission from the production or decay subprocesses of an on-shell
$V$~boson (see Eq.~(2.11) in Ref.~\cite{Dittmaier:2014qza}). This is
illustrated in Fig.~\ref{fig:amp_NNLO:propagator}, where the double
slash on a propagator line indicates that the corresponding momentum
is set on its mass shell in the rest of the diagram (but not on the
slashed line itself).  Using this decomposition, the double-real
corrections can be divided consistently into initial--initial and
initial--final contributions, as shown in
Fig.~\ref{fig:amp_NNLO:production} and Fig.~\ref{fig:amp_NNLO:decay},
respectively.  Here a diagrammatic notation is used where an encircled
diagram with an attached photon or gluon stands for all possibilities
to attach the photon/gluon to the fermion line and the gauge boson
$V$ (see Eq.~(2.12) in Ref.~\cite{Dittmaier:2014qza} for an
example). The initial--final (virtual QCD)$\times$(real EW) corrections
are treated analogously.
All different contributions to the
factorizable initial--final corrections are diagrammatically
characterized in terms of interference diagrams in
Fig.~\ref{fig:NNLO-if-graphs}.

\subsection{Calculation of the factorizable initial--final corrections}
\label{sec:calc-fact-nnlo-if}

In this section we calculate the various contributions to the
factorizable initial--final corrections of \order{\alphas\alpha} shown
in Fig.~\ref{fig:NNLO-if-graphs}.  Most contributions can be expressed in terms of reducible products
of NLO QCD and NLO EW building blocks. For details on the notation
used for these NLO results we refer to Ref.~\cite{Dittmaier:2014qza}.

\subsubsection{Double-virtual corrections}
\label{sec:IF:VV}

The double-virtual \order{\alphas\alpha} initial--final corrections to the
squared $\Paq_a\Pq_b\to\Pl_1\Pal_2$ amplitude are illustrated in
Fig.~\ref{fig:NNLO-if-graphs}(a) in terms of interference diagrams.
They arise in two ways: from the interference of the tree amplitude
with the two-loop \order{\alphas\alpha} amplitude 
and from the interference
between the one-loop amplitudes with \order{\alphas} corrections to
$V$-boson production and \order{\alpha} corrections to the decay, respectively,
\begin{align}
    \bigl\lvert \cM^{\Paq_a\Pq_b\to\Pl_1\Pal_2} \bigr\rvert^2 
    \;\Bigr\rvert_{\pro\times\dec}^{\Vs\otimes\Vew} &= 
    \hphantom{+} 2\Re\left\{
    \delta\cM_{\Vs\otimes\Vew,\pro\times\dec}^{\Paq_a\Pq_b\to\Pl_1\Pal_2}
    \left(\cM_{0,\PA}^{\Paq_a\Pq_b\to\Pl_1\Pal_2}\right)^* \right\}
    \nonumber\\&\quad
    + 2\Re\left\{
    \delta\cM_{\Vew,\dec}^{\Paq_a\Pq_b\to\Pl_1\Pal_2}
    \left(\delta\cM_{\Vs,\PA}^{\Paq_a\Pq_b\to\Pl_1\Pal_2}\right)^* \right\} .
    \label{eq:IF:NNLO-vv}
\end{align}
The LO amplitude in PA, $\M_{0,\PA}$, differs from the full LO
matrix element by the absence of the non-resonant photon diagram in
case of the neutral-current Drell--Yan process. 
 The first term on the
right-hand side in Eq.~\eqref{eq:IF:NNLO-vv} involves the factorizable
initial--final contribution to the two-loop amplitude, which takes the form
of reducible (one-loop)$\times$(one-loop) diagrams and is defined
explicitly as
\begin{equation}
  \label{eq:IF:mix_vv1}
    \delta\cM_{\Vs\otimes\Vew,\pro\times\dec}^{\Paq_a\Pq_b\to\Pl_1\Pal_2} =
  \sum_{\lambda_V} 
  \frac{
    \delta\cM_\Vs^{\Paq_a\Pq_b\to V}({\lambda_V})\;
    \delta\cM_\Vew^{V\to\Pl_1\Pal_2}({\lambda_V})
  }{p_V^2-\mu_V^2} 
= \delta_{\Vs}^{V\Paq_a\Pq_b}\; \deltadec \;  
  \cM_{0,\PA}^{\Paq_a\Pq_b\to\Pl_1\Pal_2} ,
\end{equation}
where a sum over the physical polarization states of the
vector boson $V$, labelled by $\lambda_V$, is performed. In the second
step in Eq.~\eqref{eq:IF:mix_vv1} the fact is used that the one-loop
corrections to the production and decay factorize off the
corresponding LO matrix elements,
\begin{align}
  \label{eq:factQCD:pro:virt}
   \delta \cM_{\Vs}^{\Paq_a\Pq_b\to V} &=\delta_{\Vs}^{V \Paq_a\Pq_b}
 \;\cM_{0}^{\Paq_a\Pq_b\to V},  \\
  \label{eq:factEW:dec:virt}
  \delta \cM_{\Vew,\dec}^{V\to\Pl_1\Pal_2} &=
  \deltadec 
  \;\cM_{0,\PA}^{V\to\Pl_1\Pal_2} .
\end{align}
The virtual QCD corrections are well known and are quoted explicitly in 
Eq.~(2.35) of Ref.~\cite{Dittmaier:2014qza}.
The explicit expressions for the NLO EW correction factors can be found, 
e.g., in Refs.~\cite{Dittmaier:2001ay,Dittmaier:2009cr}, and are quoted in 
Appendix B.2 of Ref.~\cite{Huss:2014eea}.
In order to maintain gauge invariance, the NLO production 
and decay subamplitudes in Eq.~\eqref{eq:IF:mix_vv1}, 
and in particular
the correction factor $\deltadec$, are evaluated for on-shell $V$ bosons. 
We keep the QCD correction factor $\delta_{\Vs}^{V\Paq_a\Pq_b}$ off shell,
i.e.\ without setting $s\to M_V^2$ there to be closer to the full calculation,
which is possible, because $\delta_{\Vs}^{V\Paq_a\Pq_b}$ does not depend on
$M_V$ at all.
The on-shell projection $s\to M_V^2$ in the EW correction
involves some freedom, but numerical
effects from different implementations are of the same order as the
intrinsic uncertainty of the PA.  However, the choice of the mappings
in the virtual and real corrections has to match properly in order to
ensure the correct cancellation of IR singularities. 

The expressions~\eqref{eq:factQCD:pro:virt}
and~\eqref{eq:factEW:dec:virt} also enter the one-loop interference
terms in the second line of Eq.~\eqref{eq:IF:NNLO-vv}.  The final
  result for the double-virtual corrections to the cross section is
  therefore given by
\begin{align}
  \label{eq:IF:mix_vv}
  \rd\sigma_{\Paq_a\Pq_b,\pro\times\dec}^{\Vs\otimes\Vew} &= 
  2 \biggl[
  \Re\Bigl\{ \deltadec \; \delta_{\Vs}^{V\Paq_a\Pq_b} \Bigr\}
  + \Re\Bigl\{ \deltadec \Bigl(\delta_{\Vs}^{V\Paq_a\Pq_b} \Bigr)^* \Bigr\}
  \biggr] \; \rd\sigma_{\Paq_a\Pq_b,\PA}^0 
  \nonumber\\&=
  4\Re\Bigl\{ \deltadec \Bigr\}
  \Re\Bigl\{ \delta_{\Vs}^{V\Paq_a\Pq_b} \Bigr\}  \;
  \rd\sigma_{\Paq_a\Pq_b,\PA}^0 .
\end{align}
Both the EW and QCD correction factors contain soft and collinear
singularities, which take the form of $\frac{1}{\epsilon^2}$ poles for
massless fermions. Therefore, in principle, Eq.~\eqref{eq:IF:mix_vv}
requires the evaluation of the correction factors up to
\order{\epsilon^2} in order to obtain all finite \order{\epsilon^0}
terms. %
However, after applying the subtraction formalism, which we describe
in detail in Sect.~\ref{sec:IF:master}, the poles are cancelled before
performing the full expansion in $\epsilon$ and, thus, the results up
to order \order{\epsilon^0} turn out to be sufficient.  This result is obvious if
the soft and collinear singularities are not regularized in
$D=4-2\epsilon$ dimensions,
but by small mass parameters, where no rational terms 
from the multiplication of $1/\epsilon$ poles with $D$-dimensional quantities
exist at all.

\subsubsection{(Real QCD)\texorpdfstring{${\times}$}{x}(virtual EW) corrections}
\label{sec:IF:RV}

The (real QCD)$\times$(virtual EW) contributions to the factorizable initial--final corrections shown in Fig.~\ref{fig:NNLO-if-graphs}(\subref{fig:NNLO-if-graphs-rv}) arise by 
including the virtual corrections to the leptonic $\PW$/$\PZ$ decays  to the various partonic subprocesses of $V+\text{jet}$ production, 
\begin{subequations}
\label{eq:rQCD}
\begin{align}
  \Paq_a(p_a) \,+\, \Pq_b(p_b) &\,\to\, V(p_V) \,+\, \Pg(k_{\Pg}) ,\\
  \Pg(p_{\Pg}) \,+\, \Pq_b(p_b) &\,\to\, V(p_V) \,+\, \Pq_a(k_a) ,\\
  \Pg(p_{\Pg}) \,+\, \Paq_a(p_a) &\,\to\, V(p_V)  \,+\, \Paq_b(k_b) .
\end{align}
\end{subequations}
For the quark-induced channel, the corrections are given by replacing the virtual QCD amplitude in Eq.~\eqref{eq:IF:mix_vv1} by the corresponding amplitude for real-gluon emission,
\begin{align}
  \label{eq:IF:mix-rv-gen}
  \delta\cM_{\Rs\otimes\Vew,\pro\times\dec}^{\Paq_a\Pq_b\to\Pl_1\Pal_2\Pg} &=
  \sum_{\lambda_V} 
  \frac{
    \cM_{\Rs}^{\Paq_a\Pq_b\to \Pg V}({\lambda_V})\;
    \delta\cM_\Vew^{V\to\Pl_1\Pal_2}({\lambda_V})
  }{p_V^2-\mu_V^2} .
\end{align}
Analogously to the double-virtual case, the EW decay
subamplitude is evaluated for on-shell vector bosons,
while the QCD correction is kept off shell.
Using the factorization property of the EW one-loop decay corrections~\eqref{eq:factEW:dec:virt}, the (real QCD)$\times$(virtual EW) correction to the cross section in the quark--anti-quark channel is proportional to the real NLO QCD corrections $\rd\sigma^{\Rs}$,  
\begin{equation}
\label{eq:IF:mix-rv}
  \rd\sigma_{\Paq_a\Pq_b,\pro\times\dec}^{\Rs\otimes\Vew}
  = 2\Re\Bigl\{ \deltadec \Bigr\}\rd\sigma_{\Paq_a\Pq_b,\PA}^{\Rs}  .
\end{equation}
As for the Born amplitude, the label PA in the real-emission corrections indicates that all non-resonant terms, i.e.\ the photon-exchange diagrams in case of the neutral-current process, are omitted in the QCD real-emission amplitudes. 
Analogous expressions hold for the gluon--quark and gluon--anti-quark initiated subprocesses to $V+\text{jet}$ production.

\subsubsection{(Virtual QCD)\texorpdfstring{${\times}$}{x}(real photonic) corrections}
\label{sec:IF:VR}

The (virtual QCD)$\times$(real photonic) factorizable corrections of
initial--final type arise from the generic interference diagram shown
in Fig.~\ref{fig:NNLO-if-graphs}(\subref{fig:NNLO-if-graphs-vr}). They
are obtained by combining the real-photon corrections to on-shell
$V$-boson decay with the virtual QCD corrections to $V$-boson
production,
\begin{equation}
\cM_{\Vs\otimes\Rew,\pro\times\dec}^{\Paq_a\Pq_b\to\Pl_1\Pal_2\Pgg}
 =\sum_{\lambda_V} \frac{\delta \cM_\Vs^{\Paq_a\Pq_b\to V}({\lambda_V})\;
  \cM_{\Rew}^{V\to\Pl_1\Pal_2\Pgg}({\lambda_V})}{(p_V+k)^2-\mu_V^2} = 
\delta_{\Vs}^{V \Paq_a\Pq_b}\cM_{\Rew,\fact,\dec}^{\Paq_a\Pq_b\to\Pl_1\Pal_2\Pgg}.
\end{equation}
In the second step, Eq.~\eqref{eq:factQCD:pro:virt} has been used to factorize the virtual QCD correction factor from the  matrix element $ \cM_{\Rew,\fact,\dec}^{\Paq_a\Pq_b\to\Pl_1\Pal_2\Pgg}$ for the factorizable NLO decay corrections (see Eq.~(2.14) in Ref~\cite{Dittmaier:2014qza}).
Again the matrix elements for the EW 
decay subprocess is evaluated for on-shell vector bosons,
while the QCD correction factor is kept off shell. 
As illustrated in Fig.~\ref{fig:amp_NNLO}
and discussed in detail in Ref.~\cite{Dittmaier:2014qza}, the splitting of photon-emission 
effects off the intermediate $V$-boson into parts corresponding to initial- or final-state
radiation separates the two resonance propagator factors
$1/(p_V^2-\MV^2)$ and $1/[(p_V^2+k)^2-\MV^2]$, respectively, where $p_V=k_1+k_2$.
For factorizable EW decay correction we, thus, have to perform the on-shell projection
$(p_V^2+k)^2\to \MV^2$.
The resulting contribution 
of the (virtual QCD)$\times$(real photonic) corrections to the cross section therefore assumes the form
\begin{align}
  \label{eq:IF:mix_vr}
  \rd\sigma_{\Paq_a\Pq_b,\pro\times\dec}^{\Vs\otimes\Rew} &= 
  2 \Re\Bigl\{ \delta_{\Vs}^{V\Paq_a\Pq_b} \Bigr\}  \;
  \rd\sigma_{\Paq_a\Pq_b,\dec}^\Rew .
\end{align}

\subsubsection{Double-real corrections}
\label{sec:IF:RR}

The double-real emission corrections  are illustrated by interference diagrams in Fig.~\ref{fig:NNLO-if-graphs}(d) and are defined by the real-emission matrix elements for the $V+\text{jet}$ production subprocesses~\eqref{eq:rQCD} with the subsequent decay $V\to \Pl_1\Pal_2\Pgg$,
\begin{align}
  \label{eq:IF:mix-rr-gen}
 \cM_{\Rs\otimes\Rew,\pro\times\dec}^{\Paq_a\Pq_b\to\Pl_1\Pal_2\Pgg\Pg}
 =\sum_{\lambda_V} \frac{\cM_\Rs^{\Paq_a\Pq_b\to V\Pg}({\lambda_V})\;
  \cM_{\Rew}^{V\to\Pl_1\Pal_2\Pgg}({\lambda_V})}{(p_V+k)^2-\mu_V^2}, 
\end{align}
with analogous expressions for the $\Pg q$ and $\Pg\bar q$ channels.
The non-resonant contribution arising from the case $V=\Pgg$ in the neutral-current process is again not included.
Compact explicit results for the helicity amplitudes of the double-real corrections can be found in Ref.~\cite{Huss:2014eea}.
The double-real contribution to the cross section, $\rd\sigma_{\pro\times\dec}^{\Rs\otimes\Rew}$, 
is defined in terms of the square of the matrix element~\eqref{eq:IF:mix-rr-gen}
where the decay subamplitudes are evaluated for on-shell $V$~bosons. 
Due to the spin correlations of the production and decay matrix elements and
the full kinematics of the $2\to 4$ scattering process, the double-real
corrections do not factorize further into separate EW and QCD correction
factors, in contrast to the other classes of
 factorizable initial--final corrections.

\subsection{Treatment of infrared singularities for the factorizable initial--final corrections}
\label{sec:IF:master}

The NNLO \order{\alphas\alpha} contributions to the cross section due
to the factorizable initial--final corrections are
obtained by integrating the four contributions discussed in the
previous section over the respective phase spaces,
\begin{align}
  \label{eq:IF:generic-master-first}
  \sighat_{\pro\times\dec}^{\NNLO_{\rs\otimes\rew}} 
  &=
  \int_{2} \rd\sigma_{\pro\times\dec}^{\Vs\otimes\Vew}
  + \iint\limits_{2+\Pgg} \rd\sigma_{\pro\times\dec}^{\Vs\otimes\Rew}
   + \int_{3} \rd\sigma_{\pro\times\dec}^{\Rs\otimes\Vew}
  + \iint\limits_{3+\Pgg} \rd\sigma_{\pro\times\dec}^{\Rs\otimes\Rew}
  \nonumber\\&\quad
  + \int_{2} \rd\sigma_{\pro\times\dec}^{\Cs\otimes\Vew} 
  + \iint\limits_{2+\Pgg} \rd\sigma_{\pro\times\dec}^{\Cs\otimes\Rew} ,
\end{align}
where the additional QCD collinear counterterms in the last line were
introduced to absorb the collinear singularities associated with the
quarks and gluons in the initial state into the NLO PDFs.  Note that
the EW corrections are completely confined to the decay subprocess,
and consequently, there are 
no singularities from initial-state collinear
quark--photon splittings.  This allows us
to obtain the collinear
counterterms in the last line of
Eq.~\eqref{eq:IF:generic-master-first} from the customary NLO QCD
collinear counterterms $\rd\sigma^{\Cs}$~\cite{Catani:1996vz} by
replacing the LO cross sections by the appropriate real or virtual EW
decay corrections in the PA. Using the results of
Sect.~\ref{sec:calc-fact-nnlo-if} we can write
\begin{align}
  \label{eq:IF:generic-master}
  \sighat_{\pro\times\dec}^{\NNLO_{\rs\otimes\rew}} 
  &=
  \int_{2} 4\Re\Bigl\{\delta_{\Vs}^{V\Paq_a\Pq_b}\Bigr\}\;
  \Re\Bigl\{\deltadec\Bigr\}\;
  \rd\sigma_\PA^0
  + \iint\limits_{2+\Pgg} 2\Re\Bigl\{\delta_{\Vs}^{V\Paq_a\Pq_b}\Bigr\}\;
  \rd\sigma_\dec^{\Rew} 
  \nonumber\\&\quad
  + \int_{3} 2\Re\Bigl\{\deltadec\Bigr\}\; \rd\sigma_\PA^{\Rs} 
  + \iint\limits_{3+\Pgg} \rd\sigma_{\pro\times\dec}^{\Rs\otimes\Rew}
  \nonumber\\&\quad
  + \int_{2} 2\Re\Bigl\{\deltadec\Bigr\} \; \rd\sigma_\PA^{\Cs} 
  + \iint\limits_{2+\Pgg} \rd\sigma_{\pro\times\dec}^{\Cs\otimes\Rew} .
\end{align}

Applying the QCD dipole subtraction formalism~\cite{Catani:1996vz} 
 in order to cancel the IR
singularities associated with the QCD corrections,
Eq.~\eqref{eq:IF:generic-master} can be written in the following form,
\begin{align}
  \label{eq:IF:dipole-qcd-master}
  \sighat_{\pro\times\dec}^{\NNLO_{\rs\otimes\rew}} &= 
  \int_{2} 2\Re\Bigl\{\deltadec\Bigr\} \;
  \rd\sigma_\PA^0 \diptimes
  \Bigl[ 2 \Re\Bigl\{\delta_{\Vs}^{V\Paq_a\Pq_b}\Bigr\}+\CSI \Bigr]
  \nonumber\\&\quad
  + \iint\limits_{2+\Pgg}  
  \rd\sigma_\dec^\Rew \diptimes
  \Bigl[ 2 \Re\Bigl\{\delta_{\Vs}^{V\Paq_a\Pq_b}\Bigr\}+\CSI \Bigr]
  \nonumber\\&\quad
  + \int_{3} 2\Re\Bigl\{\deltadec\Bigr\} 
  \biggl\{ \rd\sigma_\PA^{\Rs} 
  - \sum_{\substack{\QCD\\\text{dipoles}}}\rd\sigma_\PA^0\diptimes\CSV \biggr\}
  \nonumber\\&\quad
  + \iint\limits_{3+\Pgg} \biggl\{
  \rd\sigma_{\pro\times\dec}^{\Rs\otimes\Rew}
  - \sum_{\substack{\QCD\\\text{dipoles}}}\rd\sigma_\dec^\Rew\diptimes\CSV \biggr\}
  \nonumber\\&\quad
  + \int_0^1\rd x\int_2
  2\Re\Bigl\{\deltadec\Bigr\} \;
  \rd\sigma_\PA^0 \diptimes(\CSK+\CSP) 
  \nonumber\\&\quad
  + \int_0^1\rd x\iint\limits_{2+\Pgg}
  \rd\sigma_\dec^\Rew\diptimes(\CSK+\CSP) .
\end{align}
The explicit expressions of the dipole operators  $\CSV$ and the insertion operators $\CSI$, 
$\CSK$, and $\CSP$ can be found in Ref.~\cite{Catani:1996vz}.
The symbol  $\otimes$ denotes possible additional helicity and colour correlations,
and it is implicitly assumed that the cross sections multiplying the dipole operators $\CSV$ are evaluated on the respective dipole-mapped phase-space point.
The explicit expressions associated with the NLO QCD corrections were given in Ref.~\cite{Dittmaier:2014qza}.

All individual integrals appearing in Eq.~\eqref{eq:IF:dipole-qcd-master} are now free of QCD singularities, but remain IR divergent owing to the singularities contained in the EW corrections which still need to be cancelled between the virtual corrections and the corresponding real-photon-emission parts.
For this purpose we employ the dipole subtraction formalism for photon radiation~\cite{Dittmaier:1999mb,Dittmaier:2008md}, in particular  the extension of the formalism to treat decay kinematics described in detail in Ref.~\cite{Basso:2015gca}.
As a result, we are able to arrange the six contributions in Eq.~\eqref{eq:IF:dipole-qcd-master} into a form where all IR divergences are cancelled in the integrands explicitly,
\begin{align}
   \sighat_{\pro\times\dec}^{\NNLO_{\rs\otimes\rew}}
  &=\sigreg_{\pro\times\dec}^{\Vs\otimes\Vew}
  +\sigreg_{\pro\times\dec}^{\Vs\otimes\Rew}
  +\sigreg_{\pro\times\dec}^{\Rs\otimes\Vew}
  +\sigreg_{\pro\times\dec}^{\Rs\otimes\Rew}
  +\sigreg_{\pro\times\dec}^{\Cs\otimes\Vew}
  +\sigreg_{\pro\times\dec}^{\Cs\otimes\Rew} ,
 \label{eq:IF:final-master}
\end{align}
where each term is an IR-finite object and its phase-space integration can be performed numerically in four dimensions.
Equation~\eqref{eq:IF:final-master} is our master formula for the numerical evaluation discussed in Sect.~\ref{sec:IF:numerics}.
Explicit expressions for all contributions for the quark--anti-quark
and quark--gluon induced channels are given in
Appendix~\ref{app:ir-safe}.

The first two terms in Eq.~\eqref{eq:IF:final-master} arise from the sum of the double-virtual and the (virtual QCD)$\times$(real photonic) corrections, including the insertion operators from the QCD dipole formalism, and correspond to the sum of the first two lines in Eq.~\eqref{eq:IF:dipole-qcd-master}.
Applying the dipole formalism to rearrange the IR singularities of photonic origin between the virtual and real EW corrections, we obtain the following expressions for the IR-finite virtual QCD contributions to the cross section, 
\begin{align} 
  \label{eq:IF:master-VV}
  \sigreg_{\pro\times\dec}^{\Vs\otimes\Vew}&= 
  \int_{2} 
  \Bigl[ 2\Re\Bigl\{\deltadec\Bigr\} + \Iew \Bigr] \;
  \rd\sigma_\PA^0 \diptimes
  \Bigl[ 2 \Re\Bigl\{\delta_{\Vs}^{V\Paq_a\Pq_b}\Bigr\}+\CSI \Bigr] , \\
  \label{eq:IF:master-VR}
  \sigreg_{\pro\times\dec}^{\Vs\otimes\Rew}&=
  \iint\limits_{2+\Pgg}  \biggl\{ \rd\sigma_\dec^\Rew 
  -\sum_{\substack{I,J\\I\ne J}} \rd\sigma_\PA^0 \diptimes \dVew[IJ] \biggr\}
  \diptimes
  \Bigl[ 2 \Re\Bigl\{\delta_{\Vs}^{V\Paq_a\Pq_b}\Bigr\}+\CSI \Bigr] ,
\end{align}
where the sum over the emitter--spectator pairs ($I,J$) in Eq.~\eqref{eq:IF:master-VR} extends over all particles of the decay subprocess, i.e.\ $I,J=\Pl_1,\Pal_2,V$.
We have introduced a compact notation for the QED dipoles,
\begin{align}
  \label{eq:definition:dVew}
  \dVew[IJ] &= 4\pi\alpha\; \eta_IQ_I\,\eta_JQ_J\,
  \begin{cases}
  \dsub{IV} , & \text{for } (I=\Pl_1,\Pal_2) \land (J=V) , \\
  \gsub{IJ} , & \text{for } (I=\Pl_1,\Pal_2) \land (J=\Pl_1,\Pal_2) , \\
  0 , & \text{for } I=V ,
  \end{cases}
\end{align}
where $\eta_i=1$ for incoming particles and outgoing antiparticles and $\eta_i=-1$ for incoming antiparticles and outgoing particles. 
The corresponding endpoint contributions are given by
\begin{align}
  \label{eq:nlo:dec:master:Iew}
  \Iew
  &=
  \frac{\alpha}{2\pi}\, Q_{\Pl_1}\biggl[
  (Q_{\Pl_1}-Q_{\Pl_2})\; \Dsub{\Pl_1 V}
  +Q_{\Pl_2}\; \Gsub{\Pl_1\Pal_2} 
  \biggr]
  +(\Pl_1\leftrightarrow\Pal_2) ,
\end{align}
where the functions $\gsub{}$ and $\Gsub{}$ are given in Ref.~\cite{Dittmaier:1999mb}, while  $\dsub{}$ and $\Dsub{}$ are the decay dipoles and their integrated counterparts constructed in Ref.~\cite{Basso:2015gca}.
Whenever we write $\Pl_1\leftrightarrow\Pal_2$, this implies the interchange
$Q_{\Pl_1}\leftrightarrow Q_{\Pl_2}$ of the electric charges of the respective fermions,
irrespective of their particle or antiparticle nature.

As anticipated in Sect.~\ref{sec:IF:VV}, all IR singularities contained in $\delta_{\Vs}^{V\Paq_a\Pq_b}$ cancel exactly against the corresponding poles of the $\CSI$ operator within the second square bracket of Eq.~\eqref{eq:IF:master-VV}.
Similarly, all singularities in $\deltadec$ cancel against the corresponding poles in $\Iew$ in the first square bracket of Eq.~\eqref{eq:IF:master-VV}.
As a consequence, it is sufficient to use the correction factors $\deltadec$ and $\delta_{\Vs}^{V\Paq_a\Pq_b}$ up to \order{\epsilon^0}.
Furthermore, we recall that the correction factors $\deltadec$ 
are evaluated at the on-shell point $p_V^2=\MV^2$ and, thus, are
independent of the phase-space kinematics.%

The contributions involving real QCD corrections are given by the third and forth term in Eq.~\eqref{eq:IF:final-master}.
They are obtained by applying the QED dipole subtraction formalism to the sum of the third and forth line of Eq.~\eqref{eq:IF:dipole-qcd-master} and result in the following expressions for the IR-finite real-gluon contributions to the cross section,
\begin{align}
  \label{eq:IF:master-RV}
  \sigreg_{\pro\times\dec}^{\Rs\otimes\Vew}&= 
  \int_{3} 
  \Bigl[ 2\Re\Bigl\{\deltadec\Bigr\} + \Iew \Bigr]
  \biggl\{ \rd\sigma_\PA^{\Rs} 
  - \sum_{\substack{\QCD\\\text{dipoles}}}\rd\sigma_\PA^0\diptimes\CSV \biggr\} ,
  \\
  \label{eq:IF:master-RR}
  \sigreg_{\pro\times\dec}^{\Rs\otimes\Rew}&=
  \iint\limits_{3+\Pgg} \biggl\{
  \rd\sigma_{\pro\times\dec}^{\Rs\otimes\Rew}
  - \sum_{\substack{\QCD\\\text{dipoles}}} \rd\sigma_\dec^\Rew\diptimes\CSV 
  - \sum_{\substack{I,J\\I\ne J}}\rd\sigma_\PA^\Rs \diptimes \dVew[IJ]
  \nonumber\\&\qquad
  +\sum_{\substack{\QCD\\\text{dipoles}}}
  \sum_{\substack{I,J\\I\ne J}}
  \rd\sigma_\PA^0 \diptimes\CSV \diptimes \dVew[IJ]
  \biggr\}  .
\end{align}

It is instructive to  examine the local cancellation of the IR singularities in Eq.~\eqref{eq:IF:master-RR} in more detail.
The second term inside the curly brackets of Eq.~\eqref{eq:IF:master-RR} acts as a local counterterm to the double-real emission cross section $\rd\sigma^{\Rs\otimes\Rew}$ in all regions of phase space where the additional QCD radiation becomes unresolved, i.e.\ soft and/or collinear to the beam. 
The third term inside the curly brackets of Eq.~\eqref{eq:IF:master-RR} analogously  ensures the cancellation of IR singularities in the phase-space regions where the photon becomes soft and/or collinear to a final-state lepton.
A subtlety arises in the double-unresolved cases, where the cross sections $\rd\sigma_\dec^\Rew$ and $\rd\sigma_\PA^\Rs$ become singular as well, and both subtraction terms above will simultaneously act as a local counterterm, leading to the twofold subtraction of the IR singularities.
This disparity in the double-unresolved limits is exactly compensated by the last term inside the curly brackets of Eq.~\eqref{eq:IF:master-RR}, which therefore has the opposite sign.
Note that the evaluation of this last term  involves the successive application of two dipole phase-space mappings.
Owing to the property of the factorizable initial--final corrections where the emissions in the production and decay stages of the $V$ boson proceed independently, the two dipole mappings do not interfere with each other and
the order in which they are applied is irrelevant.
A related property is the factorization of the dipole phase space, where the two one-particle subspaces associated with the two unresolved emissions can be isolated simultaneously.  
This has the important consequence
that the analytic integration over the 
gluon and photon momenta can be carried out in the same manner as  at
NLO, which allows 
us to reuse the known results for the integrated
dipoles without modification.

Finally, we consider the convolution terms with additional virtual or real EW corrections given by the last two terms in Eq.~\eqref{eq:IF:final-master}.
Since these contributions are essentially given by the lower-order (in \alphas) cross sections, convoluted with the insertion operators $\CSK$ and $\CSP$, they pose no additional complications, and the resulting IR-finite contributions to
the cross section can be written as
\begin{align}
 \label{eq:IF:master-CV}
 \sigreg_{\pro\times\dec}^{\Cs\otimes\Vew}&=
 \int_0^1\rd x\int_2
 \Bigl[ 2\Re\Bigl\{\deltadec\Bigr\} + \Iew \Bigr]\;
 \rd\sigma_\PA^0 \diptimes(\CSK+\CSP)  , \\
 \label{eq:IF:master-CR}
 \sigreg_{\pro\times\dec}^{\Cs\otimes\Rew} &= 
 \int_0^1\rd x\iint\limits_{2+\Pgg}
 \biggl\{ \rd\sigma_\dec^\Rew 
 -\sum_{\substack{I,J\\I\ne J}} 
 \rd\sigma_\PA^0 \diptimes \dVew[IJ] \biggr\}
 \diptimes(\CSK+\CSP) .
\end{align}
Owing to the Lorentz invariance of the dipole formalism, no special treatment is required in contrast to our calculation of the non-factorizable corrections discussed in Ref.~\cite{Dittmaier:2014qza}, which was carried out with the slicing method to isolate soft-photon singularities.

The results presented so far are appropriate for the case of IR-safe observables, i.e.\ for the case where
collinear photons and leptons are recombined to a ``dressed'' lepton carrying their total momentum.
For non-collinear-safe observables with respect to the final-state leptons, i.e.\ the treatment of bare muons without photon recombination, we use the method of Ref.~\cite{Dittmaier:2008md} and its extension to decay kinematics described in Ref.~\cite{Basso:2015gca}.
The required modifications are described 
in Appendix~\ref{app:non-collinear-safe}.

\subsection{Factorizable final--final corrections}
\label{sec:calc-fact-nnlo-ff}

The factorizable  NNLO corrections of final--final type arise purely from the counterterms to 
the $V\Pl_1\Pal_2$ vertex and therefore factorize from the LO matrix element,
\begin{equation}
  \label{eq:FF:mix_vv1}
  \delta\cM_{\Vs\otimes\Vew,\dec\times\dec}^{\Paq_a\Pq_b\to\Pl_1\Pal_2}
  = \delta^{\mathrm{ct}, (\alphas\alpha)}_{V \Pl_1\Pal_2}\;  \;  
  \cM_{0,\PA}^{\Paq_a\Pq_b\to\Pl_1\Pal_2} .
\end{equation}
The counterterms for the leptonic vector-boson decay only receive
contributions from the vector-boson self-energies at
$\order{\alphas\alpha}$~\cite{Chang:1981qq,Djouadi:1987gn,Djouadi:1987di,Kniehl:1988ie,Kniehl:1989yc,Djouadi:1993ss},
which enter the counterterms through the vector-boson wave-function
renormalization constants and through the renormalization constants of the
electromagnetic coupling and the weak-mixing angle. 
There is only one type of
contribution from one-loop diagrams with insertions of one-loop
$\order{\alphas}$ or $\order{\alpha}$ counterterms.
It results from massive quark loops in the vector-boson self-energies 
where the QCD mass renormalization
constant has to be taken into account. 
We make use of the expressions
for the vector-boson self-energies of Ref.~\cite{Djouadi:1993ss},
which include the QCD quark mass counterterm in the on-shell scheme.  
The expressions  for the self-energies in terms of the scalar integrals computed in Ref.~\cite{Djouadi:1993ss} are given in Appendix~\ref{app:ff}.

The vertex counterterms in the on-shell renormalization scheme are
obtained from the expressions for the corresponding NLO EW
counterterms~\cite{Denner:1991kt} upon replacing the one-loop
vector-boson self-energies by the two-loop
$\mathcal{O}(\alphas\alpha)$ results and dropping lepton wave-function
renormalization constants, which receive no correction at this
order.   
We employ the $G_\mu$ input-parameter scheme where the
 electromagnetic
 coupling constant
is derived from the Fermi constant $G_\mu$ via the relation
\begin{equation}
  \label{eq:G_mu-scheme}
  \alpha_{G_\mu} \;=\; \frac{\sqrt{2}}{\pi} G_\mu \MW^2 
  \left( 1-\frac{\MW^2}{\MZ^2} \right) .
\end{equation}
The counterterm $\delta Z_e$ for
the electromagnetic charge in the $G_\mu$ scheme  is related to the one
in the $\alpha(0)$ input-parameter scheme as follows,
\begin{equation}
\label{eq:dZeGmu}
   \delta Z_e^{G_\mu}= \delta Z_e^{\alpha(0)}
-\frac{1}{2}\Delta r .
\end{equation}
The quantity $\Delta r$ comprises all higher-order corrections to muon
decay excluding the contributions that constitute QED corrections in
the Fermi model, which are included in the definition of the muon
decay constant $G_\mu$~\cite{Sirlin:1980nh},
\begin{align}
\Delta r &=
\left.\frac{\partial\Sigma^{AA}_{\rT}(k^2)}{\partial k^2}\right\vert_{k^2=0} 
- 2\frac{\delta \sw}{\sw}+
2\frac{\cw}{\sw}\;\frac{\Sigma_{\rT}^{AZ}(0)}{\MZ^2} +
\frac{\Sigma^{WW}_{\rT}(0)- \Re\, \Sigma^{WW}_{\rT}(\MW^2)}{\MW^2} +\delta r ,
\end{align}
with the renormalization constant $\delta \sw$ 
of the weak-mixing angle and the transverse parts of the vector-boson self-energies,
$\Sigma^{VV'}_{\rT}$.  The $\mathcal{O}(\alphas\alpha)$
contribution to $\Delta r$ simplifies due to the fact that there is no contribution
to the finite remainder
$\delta r$ at this order and the photon--$\PZ$-boson mixing
self-energy $\Sigma_\rT^{AZ}$ vanishes at zero
momentum~\cite{Djouadi:1993ss}.  
Moreover, since there are no loop corrections
to the leptonic vector-boson decay at $\mathcal{O}(\alphas\alpha)$,
the vertex counterterms are finite. The expressions for the
counterterms $\delta^{\mathrm{ct}, (\alphas\alpha)}_{V \Pl_1\Pal_2}$
in terms of vector-boson self-energies are explicitly given in
Appendix~\ref{app:ff}.

The contribution of the final--final corrections to the cross-section prediction are obtained by a simple phase-space integration over the Born kinematics,
\begin{equation}
  \sighat_{\dec\times\dec}^{\NNLO_{\rs\otimes\rew}} = 
  \int_{2} 2  \Re\Bigl\{\delta^{\mathrm{ct}, (\alphas\alpha)}_{V \Pl_1\Pal_2}\Bigr\}\;
  \rd\sigma_\PA^0 .
\end{equation}

 Using the values
of the input parameters given in Eq.~\eqref{eq:params} below, the
numerical value of the counterterm for the  $\PW\Pgnl\Pal$ vertex
is given by
\begin{equation}
  \delta_{\PW\Pgnl\Pal}^{\mathrm{ct},(\alphas\alpha)}=
  \frac{\alphas\,\alpha}{\pi^2}\times 0.93.
\end{equation}
The final--final correction to the cross section for the
charged-current cross section is therefore below the $0.1\%$ level and
phenomenologically negligible.  This can be partially attributed to the choice of the $G_\mu$-scheme where universal corrections to charged-current processes are absorbed in the value of $\alpha_{G_\mu}$.
The numerical values of the counterterms $\delta^{\mathrm{ct},\tau,(\alphas\alpha)}_{\PZ\Pl\Pal}$
for the $\PZ\Pl\Pal$ vertices with lepton chiralities $\tau=\pm$
are somewhat larger, but of opposite sign:
\begin{subequations}
\begin{align}
   \delta^{\mathrm{ct},+,(\alphas\alpha)}_{\PZ\Pl\Pal}& =
  \frac{\alphas\,\alpha}{\pi^2}\times (-49.3), \\
   \delta^{\mathrm{ct},-,(\alphas\alpha)}_{\PZ\Pl\Pal}&  =
  \frac{\alphas\,\alpha}{\pi^2}\times (+31.8).
\end{align}
\end{subequations}
The resulting corrections to the neutral-current Drell--Yan
process are, however, suppressed
far below the $0.1\%$ level due to cancellations between the right-
and left-handed production channels and are therefore also negligible for all phenomenological purposes.


\section{Numerical results}
\label{sec:IF:numerics}

In this section we present the numerical results for the 
dominant mixed QCD--EW corrections to the Drell--Yan process at the LHC for a centre-of-mass energy of $\sqrt{s}=14~\TeV$. We consider the two processes 
\begin{eqnarray}
  \label{eq:nlo-num-procs}
\nonumber
  \Pp+\Pp \;\to&\; \PWp \;&\to\; \Pgnl+\Plp +X, \\
  \Pp+\Pp \;\to&\; {\PZ} \;&\to\; \Plm+\Plp +X
\nonumber
\end{eqnarray}
with electrons or muons in the final state ($\Pl=\Pe,\Pgm$). 
We further distinguish two
alternative treatments of photon radiation: In the ``dressed-lepton''
case, collinear photon--lepton configurations are treated inclusively
using a photon-recombination procedure.  As a result, the numerical
predictions do not contain large logarithms of the lepton mass, which
can be set to zero. The dressed-lepton results are appropriate
mostly for electrons in the final state. In the ``bare-muon'' case, no such
recombination is performed, reflecting the experimental situation
which allows for the detection of isolated muons.  We perform a comparison
to naive factorization prescriptions of QCD and EW corrections, as
well as to a modelling of photonic final-state radiation~(FSR) by
structure functions or a photon shower. Moreover, we estimate the impact of the NNLO QCD--EW
corrections on the measurement of the $\PW$-boson mass.

\subsection{Input parameters and event selection}
\label{sec:input-cuts}

The setup for the calculation is analogous to the one used in
Ref.~\cite{Dittmaier:2014qza}.  
The choice of input parameters closely follows
Ref.~\cite{Beringer:1900zz},
\begin{equation}
\label{eq:params}
\begin{aligned}
  \MW^\OS \;=&\; 80.385~\GeV ,
  &\GW^\OS \;=&\; 2.085~\GeV , \\
  \MZ^\OS \;=&\; 91.1876~\GeV ,
  &\Gamma_\PZ^\OS \;=&\; 2.4952~\GeV , \\
  M_\PH \;=&\; 125.9~\GeV , 
  &m_\Pqt \;=&\; 173.07~\GeV , \\
  G_\mu \;=&\; 1.1663787\times 10^{-5} ~\GeV^{-2} , 
  &\alpha(0) \;=&\; 1/137.035999074 , \\
  \alphas(\MZ) \;=&\; 0.119 .
\end{aligned}
\end{equation}
 We convert the on-shell masses
and decay widths of the vector bosons to the corresponding pole masses
and widths as spelled out in Ref.~\cite{Dittmaier:2014qza}.

The electromagnetic coupling constant used in the LO predictions is
obtained from the Fermi constant by Eq.~\eqref{eq:G_mu-scheme}.
In the charged-current process, all relative electroweak corrections are computed using $\alpha_{G_\mu}$.
In the neutral-current process, however, we follow  Ref.~\cite{Dittmaier:2009cr} and use $\alpha(0)$  consistently in the relative photonic corrections while the remaining relative weak corrections are proportional to $\alpha_{G_\mu}$.
The same prescription is applied to the relative $\order{\alpha_s\alpha}$ corrections.

The masses of the light quark flavours (\Pqu, \Pqd, \Pqc, \Pqs, \Pqb) and of the leptons are neglected throughout, with the only exception in case of non-collinear-safe observables, where the final-state collinear singularity is regularized by the mass of the muon,
\begin{equation}
  \label{eq:muon-mass}
  m_{\Pgm} \;=\; \SI{105.658369}{\MeV} .
\end{equation}
The CKM matrix is chosen diagonal in the third generation and the mixing between the first two generations is parametrized by the following values for the entries of the quark-mixing matrix,
\begin{equation}
  \label{eq:ckm}
  \lvert V_{\Pqu\Pqd} \rvert \,=\, 
  \lvert V_{\Pqc\Pqs} \rvert \,=\, 0.974, \qquad
  \lvert V_{\Pqc\Pqd} \rvert \,=\, 
  \lvert V_{\Pqu\Pqs} \rvert \,=\, 0.227. 
\end{equation}
For the PDFs we consistently use the NNPDF2.3 sets~\cite{Ball:2012cx}, where the NLO and NNLO 
QCD--EW corrections are evaluated using the NNPDF2.3QED NLO set~\cite{Ball:2013hta}, which also includes \order{\alpha} corrections.
The value of the strong coupling $\alphas(\MZ)$ quoted in Eq.~\eqref{eq:params} is dictated by the choice of these PDF sets.
For the evaluation of the full NLO
EW corrections entering the naive products below, we employ the DIS factorization scheme to absorb the mass singularities into the PDFs.
The renormalization and factorization scales are set equal, with a fixed value given by the respective gauge-boson mass, 
\begin{equation}
  \label{eq:scale}
  \mur \;=\; \muf \;\equiv\; \mu \;=\; \MV ,
\end{equation}
of the process under consideration.

For the experimental identification of the Drell--Yan process we impose the following cuts on the transverse momenta and rapidities of the charged leptons,
\begin{align}
  \label{eq:cut-lep}
  p_{\rT,\Plpm} &> 25~\GeV , &
  \lvert y_{\Plpm} \rvert &< 2.5 , 
\end{align}
and an additional cut on the missing transverse energy
\begin{equation}
  \label{eq:cut-miss}
  E_\rT^\miss > 25~\GeV ,
\end{equation}
in case of the charged-current process.
For the neutral-current process we further require a cut on the invariant mass of the lepton pair,
\begin{equation}
  \label{eq:cut-mll}
  M_{\Pl\Pl} > 50~\GeV ,
\end{equation}
in order to avoid the photon pole at $M_{\Pl\Pl}\to0$.

For the dressed-lepton case, in addition, a photon recombination procedure analogous to the one used in Refs.~\cite{Dittmaier:2001ay,Dittmaier:2009cr} is applied:
\begin{enumerate}
\item Photons close to the beam with a rapidity $\lvert \eta_{\Pgg} \rvert > 3$ are treated as beam remnants and are not further considered in the event selection.
\item For the photons that pass the first step, the angular distance to the charged leptons $R_{\Plpm\Pgg}=\sqrt{(\eta_{\Plpm}-\eta_{\Pgg})^2 + (\phi_{\Plpm}-\phi_{\Pgg})^2}$ is computed, where $\phi$ denotes the azimuthal angle in the transverse plane. 
  If the distance $R_{\Plpm\Pgg}$ 
between the photon and the closest lepton is smaller than 
$0.1$, the photon is recombined with the lepton by adding the respective four-momenta, $\Plpm(k_i)+\Pgg(k)\to\Plpm(k_i+k)$.
\item Finally, the event selection cuts from Eqs.~\eqref{eq:cut-lep}--\eqref{eq:cut-mll} are applied to the resulting event kinematics.
\end{enumerate}

\subsection{Results for the dominant factorizable corrections}
\label{sec:if-results}

The NNLO QCD--EW corrections to the hadronic Drell--Yan cross section 
are dominated by
the factorizable initial--final \order{\alphas\alpha} corrections,
$\Delta\sigma_{\pro\times\dec}^{\NNLO_{\rs\otimes\rew}}$,
which are obtained
by convoluting the corresponding partonic corrections  
$\sighat_{\pro\times\dec}^{\NNLO_{\rs\otimes\rew}}$ calculated in
Section~\ref{sec:nnlo} with the PDFs.
 Our default prediction for Drell--Yan processes is then obtained by
 adding these NNLO corrections to  the sum of 
the full NLO QCD and EW corrections, 
\begin{align} 
\label{eq:def:nnlo:if}
  \sigma^{\NNLO_{\rs\otimes\rew}} &=
  \sigma^0 +\Delta\sigma^{\NLO_\rs} + \Delta\sigma^{\NLO_\rew}
  +\Delta\sigma_{\pro\times\dec}^{\NNLO_{\rs\otimes\rew}} ,
\end{align}
where all terms are consistently evaluated with the NNPDF2.3QED NLO PDFs.
The non-factorizable corrections computed in 
Ref.~\cite{Dittmaier:2014qza} were found to have a negligible impact on the cross section and are therefore not included here.
Similarly, the factorizable corrections of ``final--final'' type discussed in Sect.~\ref{sec:calc-fact-nnlo-ff} turn out to have a negligible impact on the cross-section prediction and are therefore not included in Eq.~\eqref{eq:def:nnlo:if} either.

Our result allows to validate estimates of the NNLO QCD--EW corrections based on a naive product ansatz. For this purpose, we define the naive product of the NLO QCD cross section and the relative EW corrections,
\begin{align}
  \sigma_{\text{naive fact}}^{\NNLO_{\rs\otimes\rew}} &=
  \sigma^{\NLO_\rs} (1+\delta_{\alpha})
  \nonumber\\&=
  \sigma^0 +\Delta\sigma^{\NLO_\rs} + \Delta\sigma^{\NLO_\rew}
  +\Delta\sigma^{\NLO_\rs}\;\delta_{\alpha}  ,
  \label{eq:def:nnlo:naive:fact}
\end{align}
where the relative EW corrections are defined as the ratio of the NLO EW contribution $\Delta\sigma^{\NLO_\rew}$ with respect to the LO contribution $\sigma^0$ according to
\begin{align}
  \label{eq:def:deltaew}
  \delta_{\alpha} &\equiv 
  \frac{\Delta\sigma^{\NLO_\rew}}{\sigma^0} ,
\end{align}
where both denominator and numerator are evaluated with the same NLO PDFs, so that the EW correction factors are practically independent of the PDFs.
In order to compare the factorized expression to the NNLO corrections, we define two different versions of the NLO EW corrections in Eq.~\eqref{eq:def:deltaew}: First, based on the full \order{\alpha} correction ($\delta_\alpha$), and second, based on the dominant EW final-state correction of the PA ($\delta_\alpha^\dec$).

Defining the correction factors,%
\footnote{
  Note that the correction factor $\delta_{\alphas}'$ differs from that in the standard QCD $K$~factor $K_{\mathrm{NLO}_{\mathrm{s}}}=\sigma_{\mathrm{NLO}_{\mathrm{s}}}/\sigma_{\mathrm{LO}}\equiv 1+\delta_{\alpha_s}$ due to the use of different PDF sets in the Born contributions.
  See Ref.~\cite{Dittmaier:2014koa} for further discussion.
} 
\begin{align}
  \label{eq:def:deltas'}
  \delta^{\pro\times\dec}_{\alphas\alpha} &\equiv
  \frac{\Delta\sigma_{\pro\times\dec}^{\NNLO_{\rs\otimes\rew}}}{\sigma^\LO} , &
  \delta_{\alphas}' &\equiv 
  \frac{\Delta\sigma^{\NLO_\rs}}{\sigma^\LO} ,
\end{align}
we can cast the relative difference of our best prediction~\eqref{eq:def:nnlo:if} and the 
product ansatz~\eqref{eq:def:nnlo:naive:fact} into the following form,
\begin{align}
\label{eq:diff-naive}
  \frac{\sigma^{\NNLO_{\rs\otimes\rew}}
    -\sigma_{\text{naive fact}}^{\NNLO_{\rs\otimes\rew}}}
  {\sigma^\LO}
  &=
  \delta^{\pro\times\dec}_{\alphas\alpha} - \delta_{\alphas}' \delta_\alpha ,
\end{align}
where the LO prediction $\sigma^\LO$ in the denominators is evaluated with the LO PDFs.
The difference of the relative NNLO
correction $\delta^{\pro\times\dec}_{\alphas\alpha}$ and the naive
product $\delta_{\alphas}' \delta_\alpha^{(\dec)}$ therefore allows to
assess the validity of a naive 
product ansatz.  As observed in
Sect.~\ref{sec:calc-fact-nnlo-if}, most contributions to the
factorizable initial--final corrections take the reducible form of a
product of two NLO corrections, with the exception of the double-real
emission corrections which are defined with the full kinematics of the
$2\to 4$ phase space.
Note that the double-real contributions are the only ones where the final-state leptons receive recoils from both QCD and photonic radiation, an effect that cannot be captured by naively multiplying NLO QCD and EW corrections. 
Any large deviations between
$\delta^{\pro\times\dec}_{\alphas\alpha}$ and
$\delta_{\alphas}'\delta_\alpha^{(\dec)}$ can therefore be attributed
to this type of contribution. The difference of the naive product
defined in terms of $\delta_\alpha^{\dec}$ and $\delta_\alpha$
allows us to assess the  impact of the missing $\order{\alphas\alpha}$ corrections beyond the initial--final
corrections considered in our calculation and therefore also provides
an error estimate of the PA, and in particular of the omission of the
corrections of initial--initial type.

\begin{figure}
  \centering
  \begin{minipage}[b]{.48\linewidth}
  \centering
  \includegraphics[width=\textwidth]{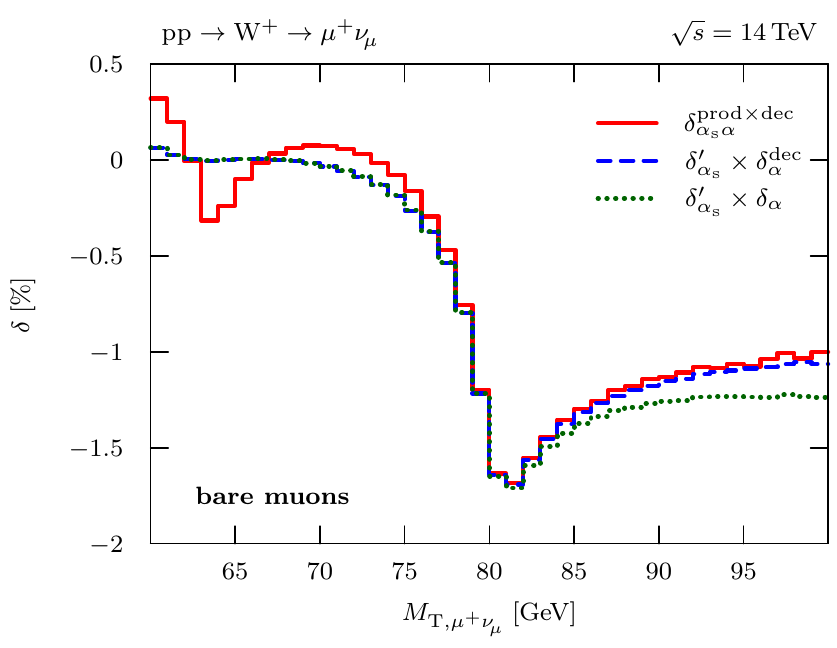} \\
  \includegraphics[width=\textwidth]{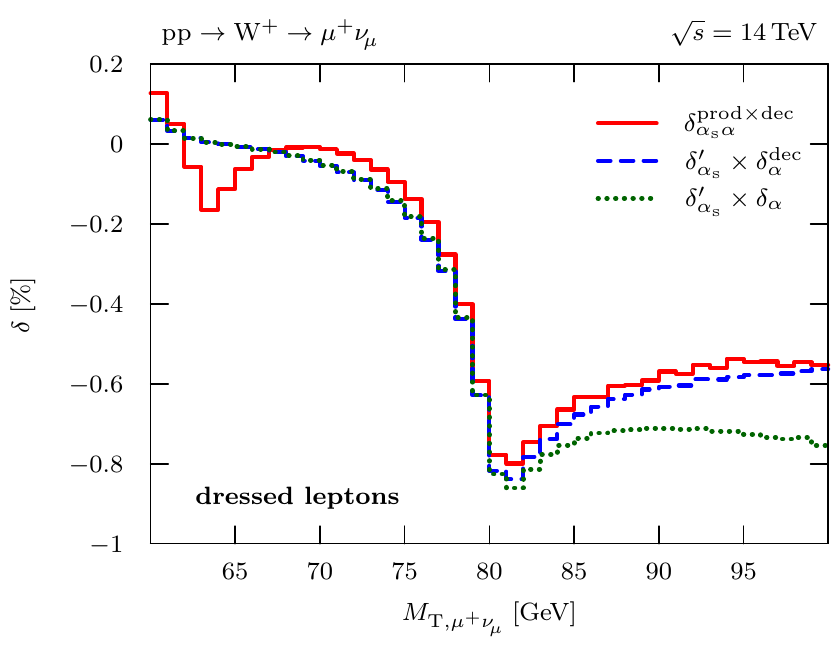}
  \end{minipage}
  \hfill
  \begin{minipage}[b]{.48\linewidth}
  \centering
  \includegraphics[width=\textwidth]{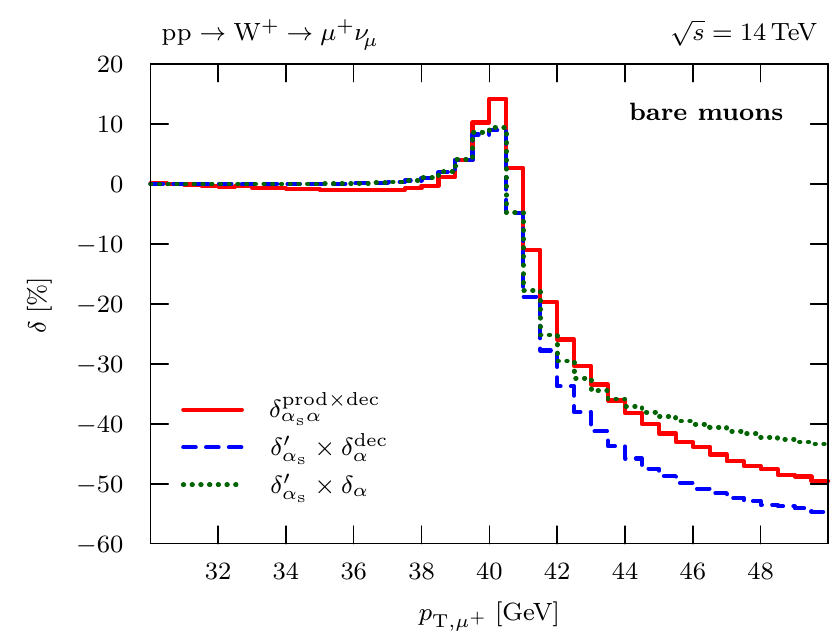} \\
  \includegraphics[width=\textwidth]{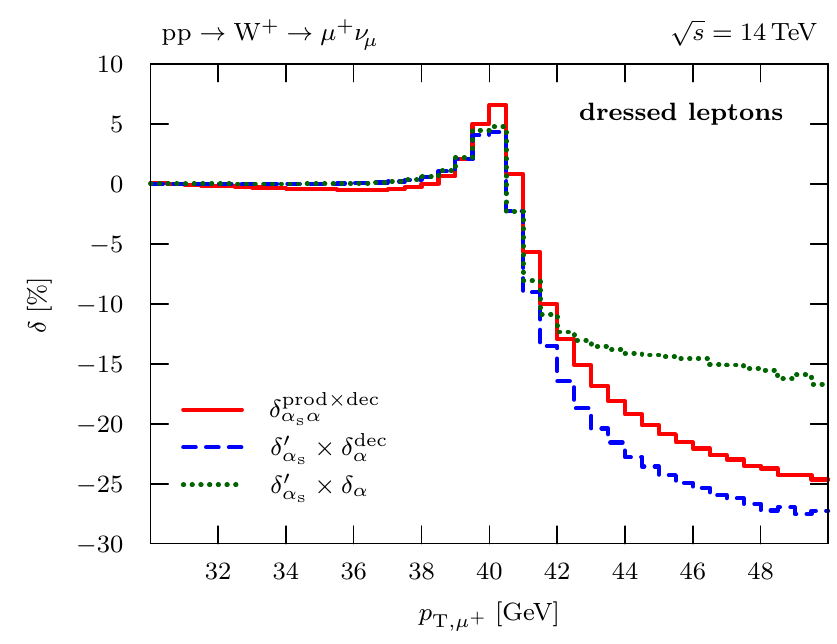}
  \end{minipage}
  \caption{Relative factorizable corrections of \order{\alphas\alpha} induced by initial-state QCD and final-state EW contributions to the transverse-mass~(left) and transverse-lepton-momentum~(right) distributions for \PWp production at the LHC.
  The naive products of the NLO correction factors $\delta_{\alphas}'$ and $\delta_\alpha$ are shown for comparison.}
  \label{fig:distNNLO-IF-Wp}
\end{figure}

\begin{figure}
  \centering
  \begin{minipage}[b]{.48\linewidth}
  \centering
  \includegraphics[width=\textwidth]{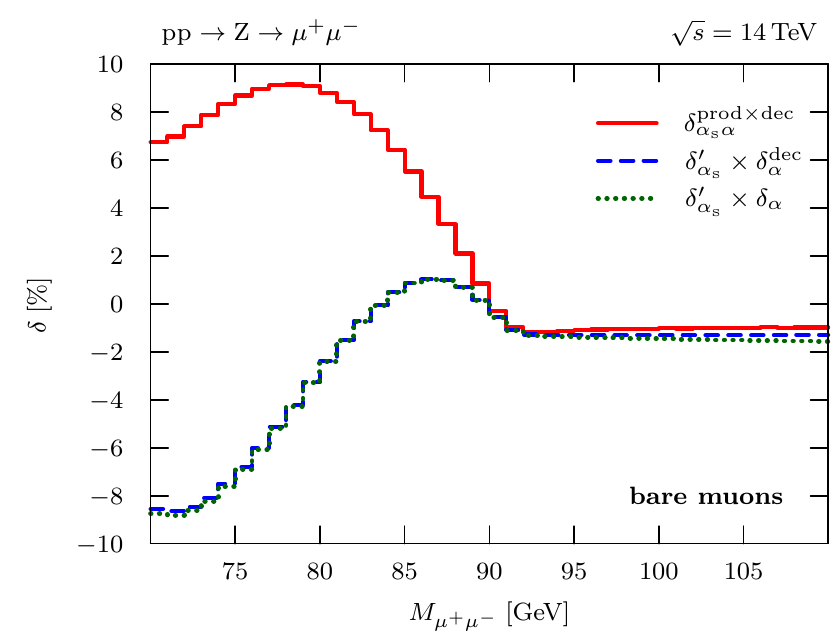} \\
  \includegraphics[width=\textwidth]{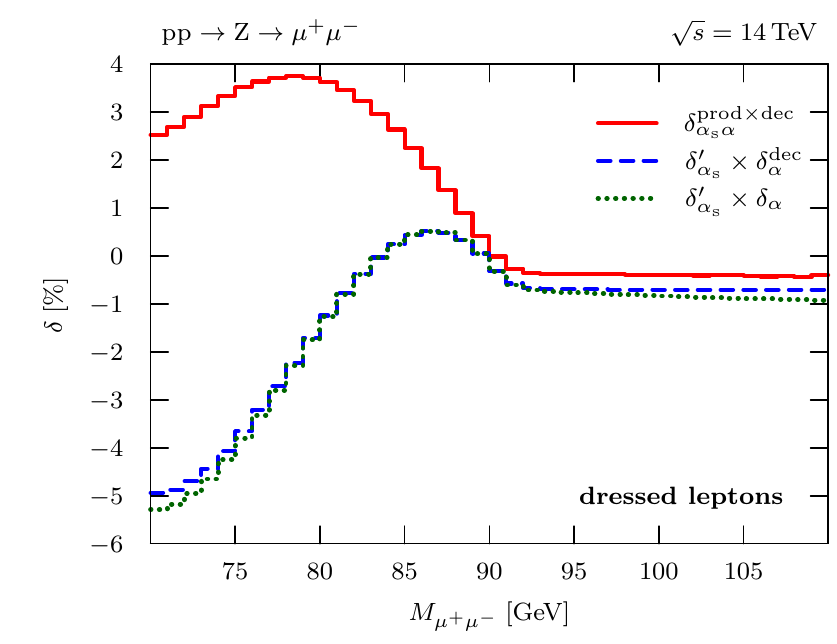}
  \end{minipage}
  \hfill
  \begin{minipage}[b]{.48\linewidth}
  \centering
  \includegraphics[width=\textwidth]{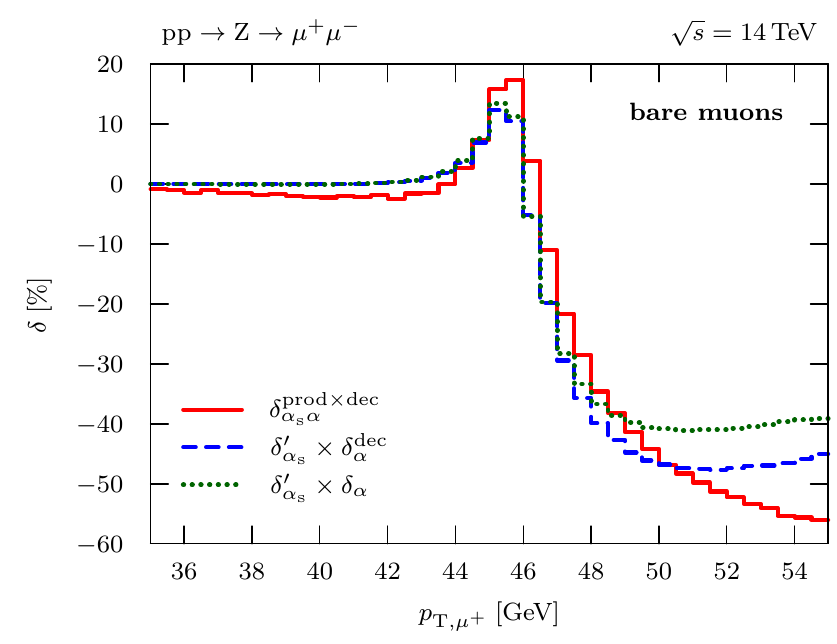} \\
  \includegraphics[width=\textwidth]{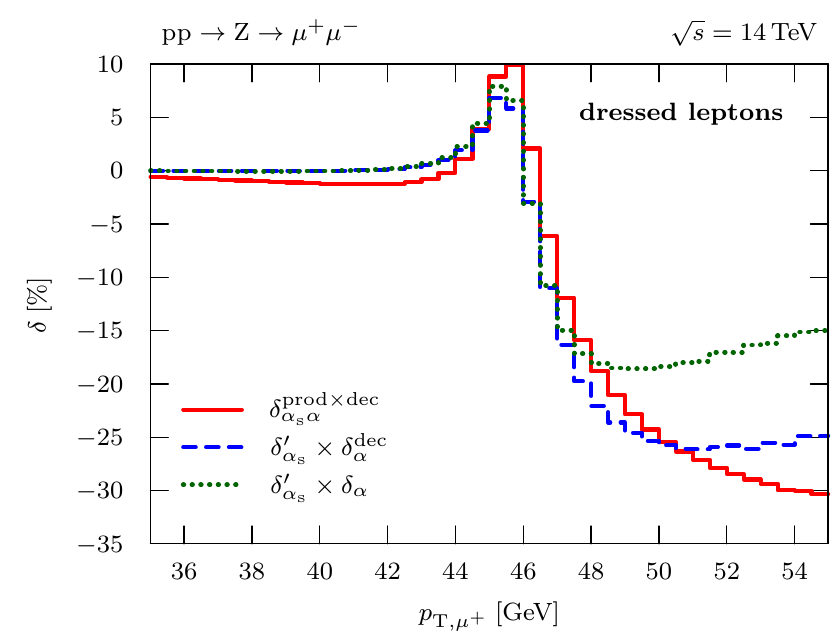}
  \end{minipage}
  \caption{Relative factorizable corrections of \order{\alphas\alpha} induced by initial-state QCD and final-state EW contributions to the lepton-invariant-mass distribution~(left) and a transverse-lepton-momentum 
distribution~(right) for \PZ production at the LHC.
  The naive products of the NLO correction factors $\delta_{\alphas}'$ and $\delta_\alpha$ are shown for comparison.}
  \label{fig:distNNLO-IF-Z}
\end{figure}

Figure~\ref{fig:distNNLO-IF-Wp} shows the numerical results for the
relative \order{\alphas\alpha} initial--final factorizable corrections $\delta^{\pro\times\dec}_{\alphas\alpha}$ to the transverse-mass ($M_{\rT,\Pgn\Pl}$) and the
transverse-lepton-momentum ($p_{\rT,\Pl}$) distributions for \PWp
production at the LHC.  For \PZ production,
Figure~\ref{fig:distNNLO-IF-Z} displays the results for the
lepton-invariant-mass ($M_{\Pl\Pl}$) distribution and a
transverse-lepton-momentum ($p_{\rT,\Pl^+}$) distribution.  In both
figures, the upper
plots show the results for bare muons, the lower panels correspond to
the corrections with photon recombination. 
In Figs.~\ref{fig:distNNLO-IF-Wp} and~\ref{fig:distNNLO-IF-Z} we
also compare to the two different implementations of a naive
product of correction factors discussed after
Eq.~\eqref{eq:diff-naive}.  
 In the following, we
mainly focus on the results for bare muons.  The respective results
with photon recombination display the same general features as those
for bare muons, but the relative corrections
are reduced by approximately a factor of two.
This reduction is familiar from NLO EW results and is induced by the
cancellation of the collinear singularities by restoring the level of
inclusiveness required for the KLN theorem.
One observes that the NNLO
$\delta^{\pro\times\dec}_{\alphas\alpha}$ corrections are in general better
approximated by the simple product ansatz for the case of bare muons
than for dressed leptons. This can
be understood from the fact that the dominant part of the corrections
stem from the collinear logarithms $\ln(m_{\Pgm})$ which are known to
factorize.

For the $M_{\rT,\Pgn\Pl}$ distribution for $\PWp$ production (upper left plot in
Fig.~\ref{fig:distNNLO-IF-Wp}), the mixed NNLO QCD--EW corrections
for bare muons are moderate and amount to approximately $\SI{-1.7}{\%}$ around the
resonance, which is about an order of magnitude smaller than the NLO
EW corrections.\footnote{The structure observed in the correction
$\delta^{\pro\times\dec}_{\alphas\alpha}$ around
$M_{\rT,\Pgn\Pl}\approx\SI{62}{\GeV}$ can be attributed to the
interplay of the kinematics of the double-real emission
corrections and the event selection. It arises close to the kinematic
boundary $M_{\rT,\Pgn\Pl}>\SI{50}{\GeV}$ for the back-to-back kinematics of the non-radiative process implied by the  cut
$p_{\rT,\Plpm},E_{\rT}^\text{miss}>\SI{25}{\GeV}$.} 
Both variants of the naive product provide a good approximation to the
full result in the region around and below the Jacobian peak, which is
dominated by resonant $\PW$ production.  For larger values of
$M_{\rT,\Pgn\Pl}$, the product $\delta_{\alphas}' \delta_\alpha $
based on the full NLO EW correction factor deviates from the other curves, which signals
the growing importance of effects beyond the PA. However, the
deviations amount to only few per-mille for
$M_{\rT,\Pgn\Pl}\lesssim\SI{90}{\GeV}$.
The overall good agreement between the
$\delta^{\pro\times\dec}_{\alphas\alpha}$ corrections and both naive
products can be attributed to  well-known insensitivity of the
observable $M_{\rT,\Pgn\Pl}$ to initial-state radiation effects
already seen in the case of NLO corrections in Ref~\cite{Dittmaier:2014qza}.
 
For the $p_{\rT,\Pl}$ distributions in the case of bare muons (upper right plots
in Figs.~\ref{fig:distNNLO-IF-Wp} and \ref{fig:distNNLO-IF-Z},
respectively) we observe corrections that are small far below the
Jacobian peak, but which rise to about $15\%$ ($20\%$) on the Jacobian
peak at $p_{\rT,\Pl}\approx\MV/2$ for the case of the \PWp boson (\PZ
boson) and then display a steep drop reaching almost $-50\%$ at
$p_{\rT,\Pl}=\SI{50}{\GeV}$.  This enhancement stems from the large
QCD corrections above the Jacobian peak familiar from the NLO QCD
results (see e.g.\ Fig.~8 in Ref.~\cite{Dittmaier:2014qza}) where
the recoil due to real
QCD radiation shifts events with resonant $\PW$/$\PZ$ bosons above the
Jacobian peak.
The
naive product ansatz fails to provide a good description of the full
result $\delta^{\pro\times\dec}_{\alphas\alpha}$ and deviates by
5--10\% at the Jacobian peak, where the PA is expected to be the most
accurate.  This can be attributed to the strong influence of the
recoil induced by initial-state radiation on the transverse momentum,
which implies a larger effect of the double-real emission corrections
on this distribution that are not captured correctly by the naive
products.  The two versions of the naive products display larger
deviations than in the $M_{\rT,\Pgn\Pl}$ distribution discussed above,
which signals a larger impact of the missing \order{\alphas\alpha}
initial--initial corrections.  However, these deviations should be
interpreted with care, 
since a fixed-order
prediction is not sufficient to describe this distribution around the
peak region $p_{\rT,\Pl}\approx\MV/2$, which corresponds to the
kinematic onset for $V+\text{jet}$ production and is known to
require QCD resummation for a proper description.

In case of the $M_{\Pl\Pl}$
distribution for $\PZ$ production (left-hand plots in
Fig.~\ref{fig:distNNLO-IF-Z}),  corrections up to $10\%$  are
observed below the resonance for the case of bare muons. This is consistent with the large EW
corrections at NLO in this region, which arise from final-state photon
radiation that shifts the reconstructed value of the invariant
lepton-pair mass away from the resonance to lower values.
The naive product approximates the
full initial--final corrections
$\delta^{\pro\times\dec}_{\alphas\alpha}$ reasonably well at the resonance
itself ($M_{\Pl\Pl}=\MZ$) and  above, but completely fails already a
little below the resonance where the naive products do not even
reproduce the sign of the full
$\delta^{\pro\times\dec}_{\alphas\alpha}$ correction.
This deviation occurs although  the invariant-mass distribution is 
widely unaffected by initial-state radiation effects. 
The fact that we obtain almost identical corrections 
from the two versions of the product $\delta_{\alphas}'\delta_\alpha^\dec$ 
and $\delta_{\alphas}'\delta_\alpha$ demonstrates the insensitivity of this
observable to photonic initial-state radiation.

\begin{figure}
  \centering
  \begin{subfigure}[b]{.48\linewidth}
  \centering
  \includegraphics[width=\textwidth]{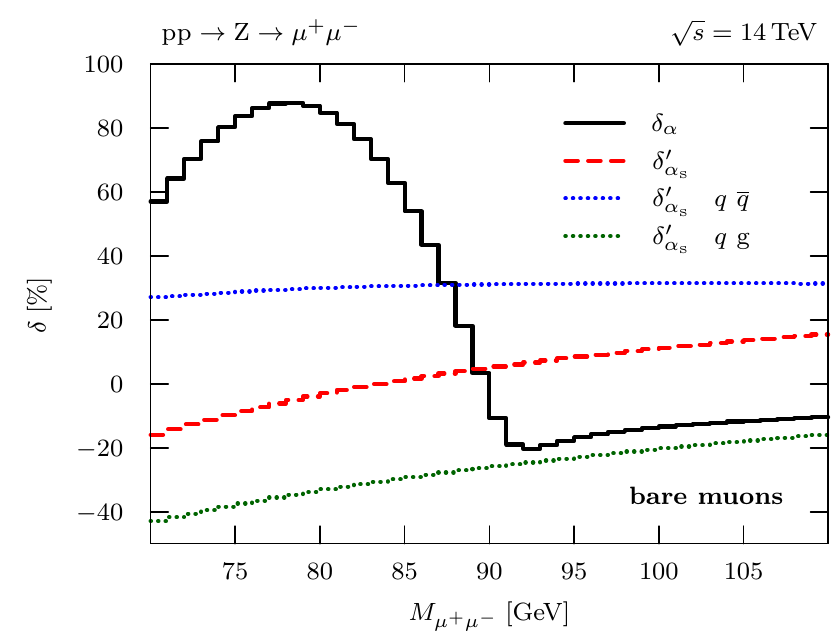}
  \caption{}
  \label{fig:distNNLO-IFZ:NLO-breakdown}
  \end{subfigure}
  \hfill
  \begin{subfigure}[b]{.48\linewidth}
  \centering
  \includegraphics[width=\textwidth]{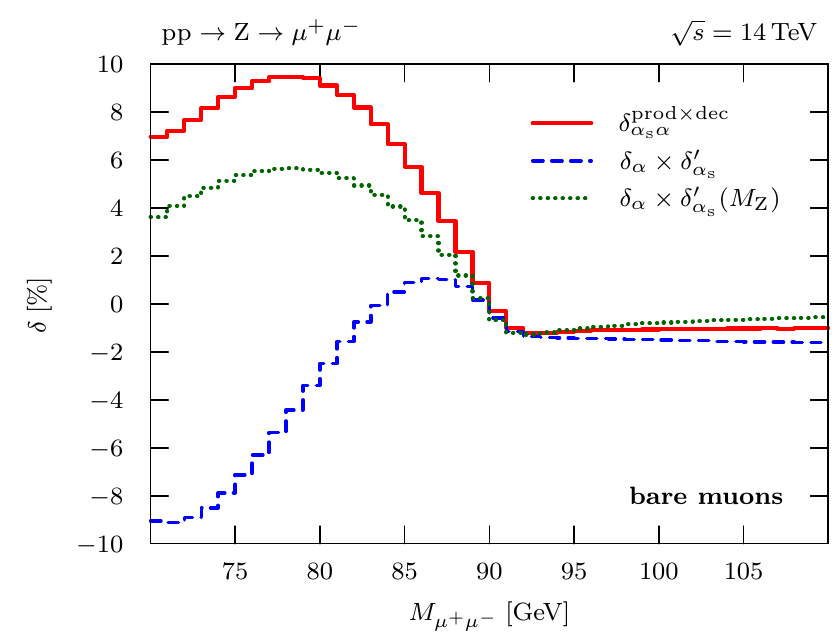}
  \caption{}
  \label{fig:distNNLO-IFZ:more}
  \end{subfigure}
  \caption{(\subref{fig:distNNLO-IFZ:NLO-breakdown})~The correction factors 
to Z-boson production (as shown in the upper left plot of Fig.\ref{fig:distNNLO-IF-Z})
entering the naive product ansatz broken down into its individual contributions with the QCD corrections further divided into the $\Pq\Paq$- and $\Pq\Pg$-induced contributions.
  (\subref{fig:distNNLO-IFZ:more})~A comparison of the corrections shown in the upper left panel of Fig.~\ref{fig:distNNLO-IF-Z} with the modified product using the value of the QCD corrections at the resonance $\delta_{\alphas}'(M_{\Pl\Pl}=\MZ)\approx 6.5\%$.}
  \label{fig:distNNLO-IFZ}
\end{figure}

In order to locate the source 
of this large discrepancy we examine
the individual correction factors in the naive product in more detail.
We restrict ourselves to the case of bare muons and  the full NLO EW
correction factor $\delta_\alpha$ for definiteness, which does not
affect our conclusions.
In Fig.~\ref{fig:distNNLO-IFZ}(\subref{fig:distNNLO-IFZ:NLO-breakdown}),
 we separately
 plot the two correction factors that enter the naive 
product $\delta_{\alphas}'\times\delta_\alpha$ and further divide the QCD corrections into the 
$\Pq\Paq$- and the $\Pq\Pg$-induced contributions.
We observe that the two different $\Pq\Paq$- and $\Pq\Pg$-induced channels individually receive large QCD corrections, however, they differ in sign, 
so that large cancellations take place in the sum $\delta_{\alphas}'$.
A small mismatch in the corrections of the individual channels can therefore quickly lead to a large effect in the QCD corrections which is then further enhanced by the large EW corrections in the product ansatz $\delta_{\alphas}' \times \delta_{\alpha}$.
Moreover,
Figure~\ref{fig:distNNLO-IFZ}(\subref{fig:distNNLO-IFZ:NLO-breakdown})
reveals that the QCD correction factor $\delta_{\alphas}'$ is
responsible for the sign change at $M_{\Pl\Pl}\approx\SI{83}{\GeV}$
which is the most striking  disagreement of the naive product ansatz
with the full factorizable initial--final corrections.
This zero crossing happens more than three widths below the resonance where the cross section is reduced by almost two orders of magnitudes compared to the resonance region and, furthermore, will be very sensitive to event selection cuts, since $\delta_{\alphas}'$ arises from the cancellation of two large corrections as we have seen above.
In Fig.~\ref{fig:distNNLO-IFZ}(\subref{fig:distNNLO-IFZ:NLO-breakdown}) we
observe the large EW corrections below the resonance mentioned above, which 
arise due to the redistribution of  events near the $\PZ$
pole to lower lepton invariant masses by final-state photon radiation.
The similar form of the factorizable NNLO initial--final
corrections indicates that they mainly stem from an analogous mechanism.
This suggests that it is more appropriate to  replace
the QCD correction factor $\delta_{\alphas}'$ in the naive product by its value at the resonance
$\delta_{\alphas}'(M_{\Pl\Pl}=\MZ)\approx 6.5\%$, which corresponds to
the location of the events that are responsible for the bulk of the
large EW corrections below the resonance. 
In contrast, the naive product ansatz simply multiplies the
corrections locally on a bin-by-bin basis. This causes a mismatch in the correction factors and fails to account for the
migration of events due to FSR.
The comparison of the previous results and the modified product is shown in Fig.~\ref{fig:distNNLO-IFZ}(\subref{fig:distNNLO-IFZ:more}) and clearly shows an improvement despite its very crude construction.

Contrary to the lepton-invariant-mass distribution, the
transverse-mass distribution is dominated by events with resonant \PW
bosons even in the range below the Jacobian peak,
$M_{\rT,\Pgn\Pl}\lesssim\MW$, so it is less sensitive to the
redistribution of events to lower $M_{\rT,\Pgn\Pl}$.
This explains why the naive product can provide a good approximation
of the full initial--final NNLO corrections.  It should be emphasized,
however, that even in the case of the $M_{\rT,\Pgn\Pl}$ distribution
any event selection criteria that deplete events with resonant \PW bosons
below the Jacobian peak will result in increased sensitivity to the effects of
FSR and can potentially lead to a failure of a naive
product ansatz. 

In conclusion, simple approximations in terms of products of
correction factors have to be used with care and require a careful
case-by-case investigation of their validity.

\subsection{Leading-logarithmic approximation for final-state photon radiation}
\label{sec:FSR}

As is evident from Figs.~\ref{fig:distNNLO-IF-Wp} and~\ref{fig:distNNLO-IF-Z}, a naive 
product of QCD and EW correction factors~\eqref{eq:def:nnlo:naive:fact} is not adequate to approximate the NNLO QCD--EW corrections for all observables.  
A promising approach to a factorized approximation for the dominant initial--final corrections can be obtained by combining the full NLO QCD corrections to vector-boson production with the leading-logarithmic~(LL) approximation for the final-state corrections.
The benefit in this approximation lies in the fact that the interplay of the recoil effects from jet and photon emission is properly taken into account.
On the other hand, the logarithmic approximation neglects certain (non-universal) finite contributions, which are, however, suppressed with respect to the dominating radiation effects.

In the structure-function approach~\cite{Kuraev:1985hb}, the
leading-logarithmic approximation of the {pho\-to\-nic} decay
corrections is
combined with the NLO QCD corrections to the production by a convolution,
\begin{align}
  \label{eq:LL1FSR}
  \Delta\sigma_{\Pp\Pp, \llog\FSR}^{\NNLO_{\rs\otimes\rew}} &=
  \int \rd\sigma^{\mathrm{NLO}_\rs}(p_1,p_2;k_1,k_2) 
  \int^1_0 \rd z_1 \, \int^1_0 \rd z_2 \,
  \;\Theta_\cut\bigl(z_1k_1\bigr)
  \;\Theta_\cut\bigl(z_2k_2\bigr)
  \nonumber\\* &\quad\times 
\biggl[\,\Gamma_{\Pl_1\Pl_1}^{\llog}(z_1,Q^2)\Gamma_{\Pl_2\Pl_2}^{\llog}(z_2,Q^2)
-\delta(1-z_1)\,\delta(1-z_2)\biggr]
  \nonumber\\ &=
  \int \rd\sigma^{\mathrm{NLO}_\rs}(p_1,p_2;k_1,k_2) \int^1_0 \rd z_1 \, \int^1_0 \rd z_2 \,
  \;\Theta_\cut\bigl(z_1k_1\bigr)
  \;\Theta_\cut\bigl(z_2k_2\bigr)
  \nonumber\\*
  &\quad \times\biggl[
 \delta(1-z_2)\,\Gamma_{\Pl_1\Pl_1}^{\llog,1}(z_1,Q^2)
  +\delta(1-z_1)\,\Gamma_{\Pl_2\Pl_2}^{\llog,1}(z_2,Q^2)
  +\order{\alpha^2} \biggr] 
  ,
\end{align}
where  $\rd\sigma^{\mathrm{NLO}_\rs}$ includes the virtual and real QCD corrections.
The step function $\Theta_{\cut}(z_i k_i)$ is equal to 1 if the event
passes the cut on the rescaled lepton momentum $z_i k_i$ and 0
otherwise.  The variables $z_i$ are the momentum fractions describing the
respective 
lepton energy loss by collinear photon emission. For the charged-current process only one of the convolutions is present.
The $\order{\alpha}$ contribution to the structure function $\Gamma_{\Pl\Pl}^{\llog}$ reads
\begin{eqnarray}
\label{eq:LLll}
  \Gamma_{\Pl\Pl}^{\llog,1}(z,Q^2) &=&
  Q_{\Pl}^2\,\frac{\beta_\Pl}{4} \left(\frac{1+z^2}{1-z}\right)_+ ,
\end{eqnarray}
where the large mass logarithm appears in the variable
\begin{equation}
\label{eq:beta_l}
\beta_\Pl = \frac{2\,\alpha}{\pi}
\left[\ln\biggl(\frac{Q^2}{\Ml^2}\biggr)-1\right]
\end{equation}
and 
$Q_{\Pl}$ denotes the relative electric charge of the lepton $\ell$. 
In order to be consistent in the comparison with our calculation as described in Sect.~\ref{sec:input-cuts}, the electromagnetic coupling constant $\alpha$ appearing in Eq.~\eqref{eq:beta_l} is set to $\alpha_{G_\mu}$ and $\alpha(0)$ for the charged-current and neutral-current processes, respectively.
The scale $Q$ is chosen as the gauge-boson mass,
\begin{equation}
\label{eq:FSR_scale}
Q = M_{V} ,
\end{equation}
and the scale-variation bands shown in the numeric results are obtained by varying the scale by a factor of two up and down from the central scale choice,
\begin{equation}
Q = \xi \cdot M_V , \qquad \xi = \tfrac{1}{2}, 1, 2 .
\label{eq:ll1fsr:variation}
\end{equation}
Since the mass logarithms cancel in observables where photon emission collinear to the final-state charged leptons is treated fully inclusively, 
the structure-function approach is only applicable to non-collinear-safe observables, i.e.\ to the bare-muon case. 

In contrast, in 
parton-shower approaches to photon radiation (see e.g.\ Refs.~\cite{Placzek:2003zg,CarloniCalame:2003ux,CarloniCalame:2005vc})
the photon momenta transverse to the lepton momentum are 
generated as well, following the differential factorization formula, so that
the method is also applicable to the case of collinear-safe observables,
i.e.\ to the dressed-lepton case. For this purpose, we have
implemented the combination of the exact NLO QCD prediction for
vector-boson production with the simulation of final-state photon radiation
using PHOTOS~\cite{Golonka:2005pn}. Since we are interested in
comparing  to the
\order{\alphas\alpha} corrections in our setup, we only generate a
single photon emission using PHOTOS and use the same scheme for
$\alpha$ as described in Sect.~\ref{sec:input-cuts}. 
Details on the specific settings within the PHOTOS parton shower are given in Appendix~\ref{app:photos}.

\begin{figure}
  \centering
  \begin{minipage}[b]{.48\linewidth}
  \centering
  \includegraphics[width=\textwidth]{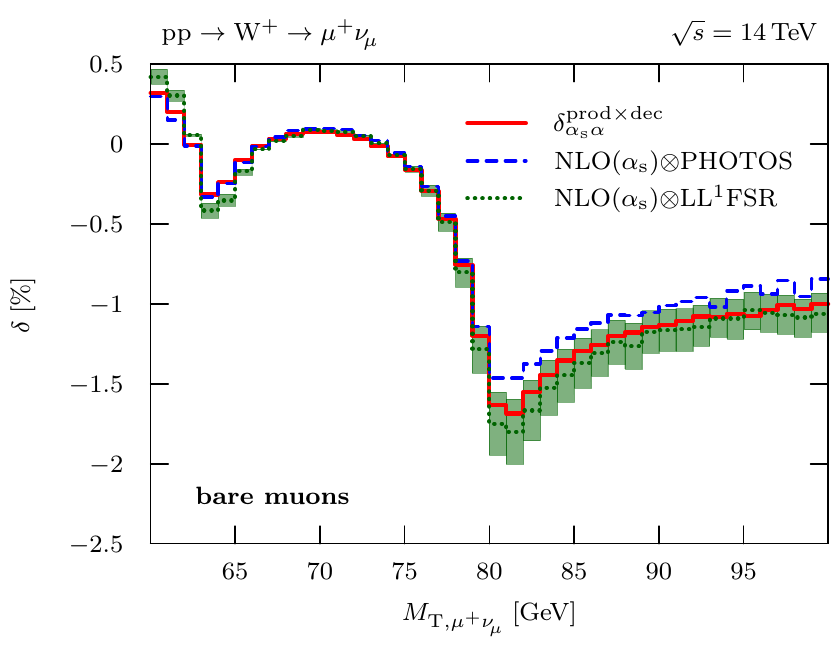} \\
  \includegraphics[width=\textwidth]{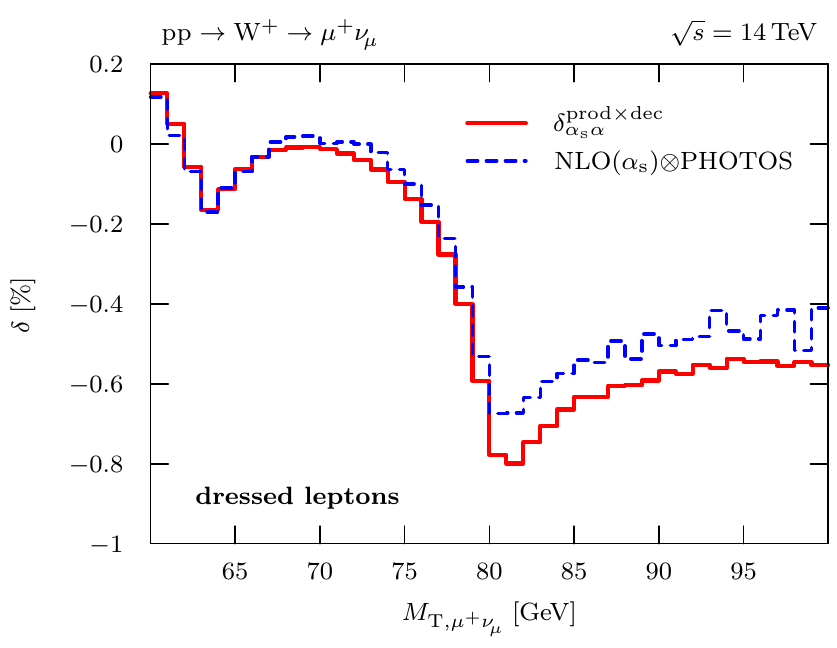}
  \end{minipage}
  \hfill
  \begin{minipage}[b]{.48\linewidth}
  \centering
  \includegraphics[width=\textwidth]{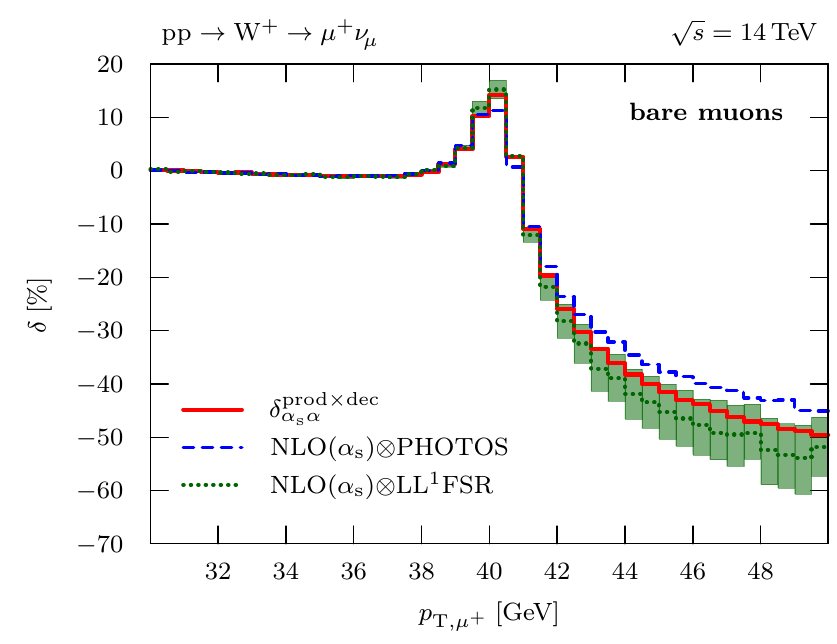} \\
  \includegraphics[width=\textwidth]{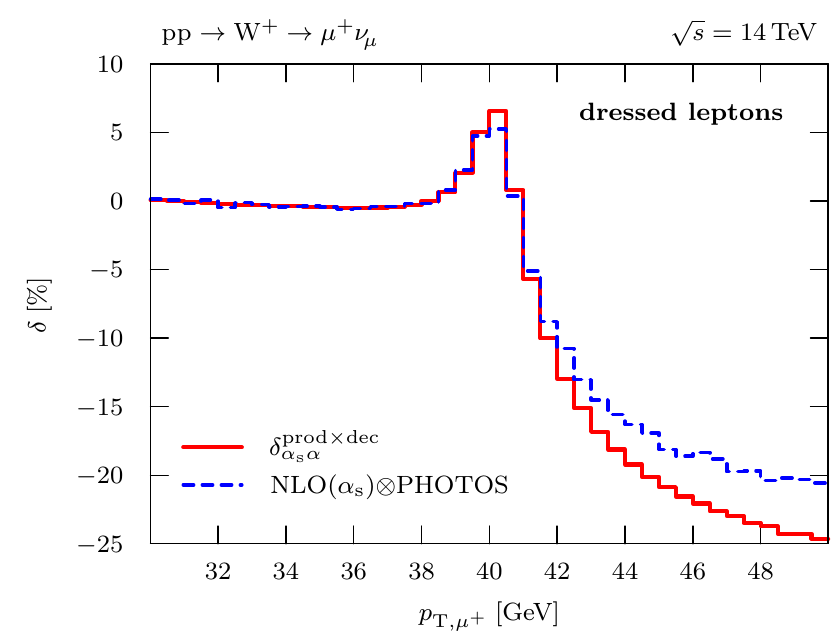}
  \end{minipage}
  \caption{Comparison of the approximation obtained
    from PHOTOS for the relative \order{\alphas\alpha} initial-state
    QCD and final-state EW corrections to our best prediction
    $\delta^{\pro\times\dec}_{\alphas\alpha}$ for the case of the
    transverse-mass~(left) and transverse-lepton-momentum~(right)
    distributions for \PWp production at the LHC, as in
    Fig.~\ref{fig:distNNLO-IF-Wp}.  In the bare-muon case, the
    result~\eqref{eq:LL1FSR} of the structure-function approach is
    also shown.  }
  \label{fig:distNNLO-IF-FSR-Wp}
\end{figure}

\begin{figure}
  \centering
  \begin{minipage}[b]{.48\linewidth}
  \centering
  \includegraphics[width=\textwidth]{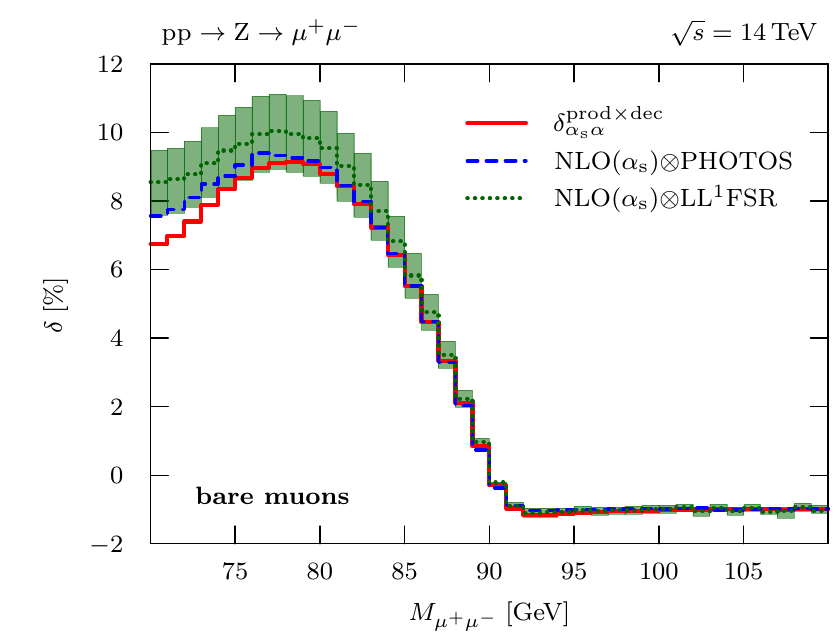} \\
  \includegraphics[width=\textwidth]{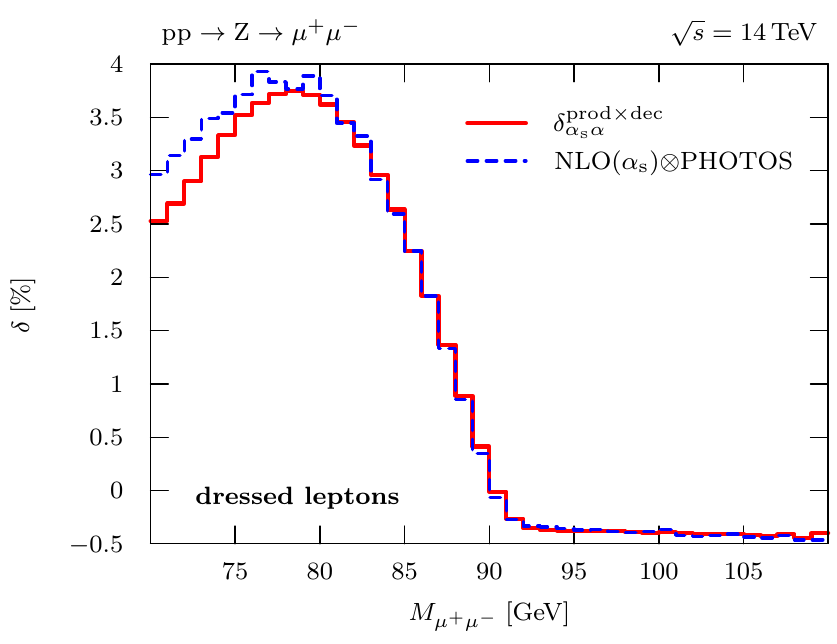}
  \end{minipage}
  \hfill
  \begin{minipage}[b]{.48\linewidth}
  \centering
  \includegraphics[width=\textwidth]{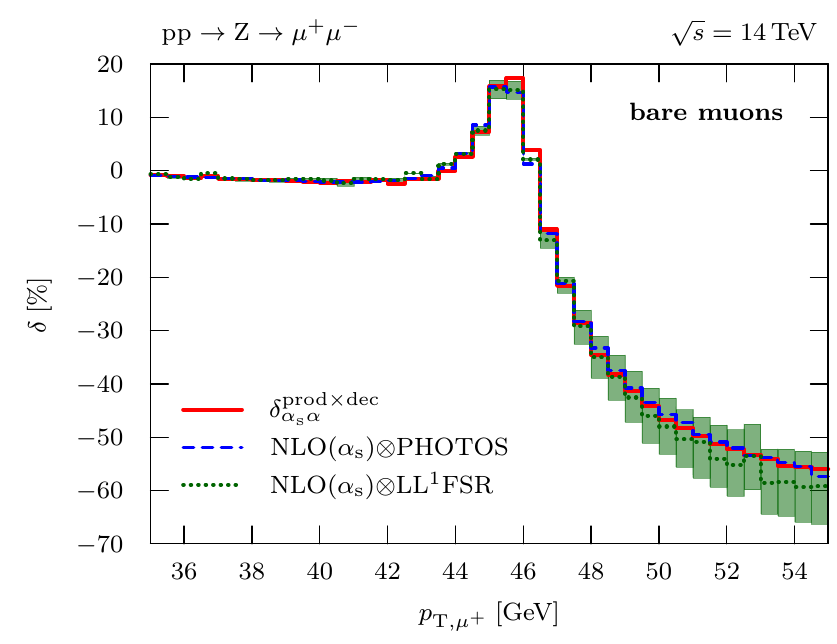} \\
  \includegraphics[width=\textwidth]{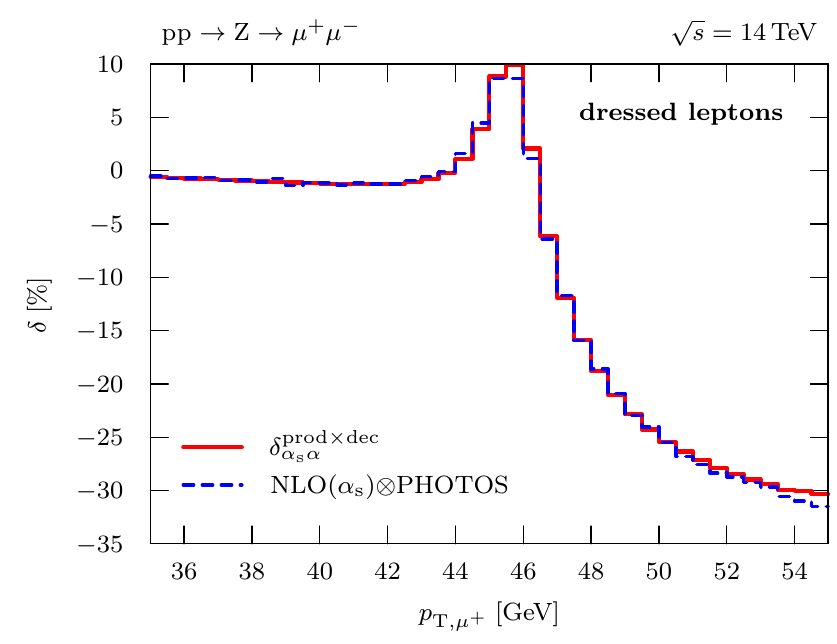}
  \end{minipage}
  \caption{Comparison of the approximation obtained from PHOTOS for the
    relative \order{\alphas\alpha} initial-state QCD and final-state
    EW corrections to our best prediction
    $\delta^{\pro\times\dec}_{\alphas\alpha}$ for the case of the
    lepton-invariant-mass distribution~(left) and a
    transverse-lepton-momentum distribution~(right) for \PZ production at
    the LHC, as in Fig.~\ref{fig:distNNLO-IF-Z}. In the bare-muon case, the
    result~\eqref{eq:LL1FSR} of the structure-function approach is
    also shown.}
  \label{fig:distNNLO-IF-FSR-Z}
\end{figure}

In Figs.~\ref{fig:distNNLO-IF-FSR-Wp} and \ref{fig:distNNLO-IF-FSR-Z}
we compare our best prediction~\eqref{eq:def:nnlo:if} for the
factorizable initial--final \order{\alphas\alpha} corrections to the
combination of NLO QCD corrections with the approximate FSR
obtained from PHOTOS for the case of $\PWp$ production and $\PZ$
production, respectively. For the bare-muon case also the result of
the structure-function approach according to Eq.~\eqref{eq:LL1FSR} is
shown. The combination of the NLO QCD corrections and 
approximate FSR leads to a clear improvement compared to the
naive product approximations investigated in
Section~\ref{sec:if-results}.  This is particularly apparent in the
neutral-current process where the $M_{\Pl\Pl}$ distribution is correctly modelled
by both FSR approximations, whereas the naive products shown in
Figs.~\ref{fig:distNNLO-IF-Z} and~\ref{fig:distNNLO-IFZ} completely failed to
describe this distribution. 
 In the $M_{\rT,\Pgn\Pl}$ spectrum of the charged-current
process in Fig.~\ref{fig:distNNLO-IF-FSR-Wp} one also finds good
agreement of the different results below the Jacobian peak and an
improvement over the naive product approximations in
Fig.~\ref{fig:distNNLO-IF-Wp}. 
The description of the $p_{\rT,\Pl}$ distributions is also improved compared to the naive product approximations, but some differences remain in the charged-current process.

In spite of the good agreement of the two versions of incorporating final-state-radiation effects, the intrinsic uncertainty of the leading-logarithmic approximations should be kept in mind.
For the structure-function approach, this uncertainty is illustrated by the band width resulting from the variation~\eqref{eq:ll1fsr:variation} of the QED scale $Q$.
We remark that the multi-photon corrections obtained by employing the
un-expanded
structure-functions $\Gamma_{\Pl\Pl}^{\llog}(z,Q^2)$  in Eq.~\eqref{eq:LL1FSR}
lie well within the aforementioned scale bands, which shows that a
proper matching to the full NLO EW calculation is needed to
remove the dominant uncertainty of the LL approximation and to 
 predict the higher-order effects reliably.
For PHOTOS the intrinsic uncertainty is not shown and not easy to
quantify.
The good quality of the PHOTOS approximation results from the fact that the finite terms in the photon emission probability are specifically adapted to $\PW/\PZ$-boson decays.
The level of agreement with our ``full prediction'', thus, cannot be taken over to other processes.

\subsection{Impact on the \texorpdfstring{$\PW$}{W}-boson mass extraction}
\label{sec:mw}

In order to estimate the effect of the \order{\alphas \alpha}
corrections on the extraction of the $\PW$-boson mass at the LHC we
have performed a $\chi^2$ fit of the $M_{\rT,\nu \ell}$ distribution.
We treat the $M_{\rT,\nu \ell}$ spectra calculated in various
theoretical approximations for a reference mass $\MW^\OS= 80.385~\GeV$
as ``pseudo-data'' that we fit with 
 ``templates'' calculated
using the LO predictions $\sigma^0$ (with NLO PDFs)
for different values of $\MW^\OS$.
Specifically, we have generated results for $27$ transverse-mass bins in the
interval $M_{\rT,\nu \ell}= [64,\,91]~\GeV$ in steps of $1~\GeV$, varying the $\PW$-boson mass in the interval $\MW=[80.085,\,80.785]~\GeV$ with steps of $\Delta \MW=10~ \MeV$ (steps of $\Delta
\MW=5~\MeV$ in the interval $\MW=[80.285,\,80.485]~\GeV$).  
Using a linear interpolation between neighbouring $\MW$ values, we obtain the integrated cross sections in the $i$th $M_{\rT,\nu \ell}$ bin, $\sigma_i^0(\MW)$, as a continuous function of $\MW$. 
The best-fit value $\MW^{\mathrm{fit}, \mathrm{th}}$ 
quantifying the impact of a higher-order correction
in the theoretical cross section $\sigma^{\mathrm{th}}$ is
then obtained from the minimum of the function
\begin{equation}
\label{eq:chi2}
  \chi^2(\MW^{\mathrm{fit},\mathrm{th}})=\sum_{i}
  \frac{\left[\sigma_i^{\mathrm{th}}(\MW^\OS)
      -\sigma_i^0(\MW^{\mathrm{fit},\mathrm{th}})\right]^2}{
    2 \Delta\sigma_i^2}\,,
\end{equation}
where the sum over $i$ runs over the transverse-mass bins.
Here $\sigma_i^{\mathrm{th}}$ and $\sigma_i^0$ are the
integrated cross sections in the $i$-th bin, uniformly rescaled so
that the sum over all $27$ bins is identical for all considered cross
sections.  We assume a statistical error of the pseudo-data and take
$\Delta\sigma_i^2\propto \sigma_i^{\mathrm{th}}$.  We have also
performed a two-parameter fit where the normalization of the templates
is fitted simultaneously, leading to identical results.  Similarly,
allowing the $\PW$-boson width in the templates to float and fitting
$\MW$ and $\GW$ simultaneously does not
significantly affect the estimate of the effect of the
\order{\alphas\alpha} corrections on the \MW measurement.

 \begin{figure}
   \centering
   \includegraphics[width=.48\linewidth]{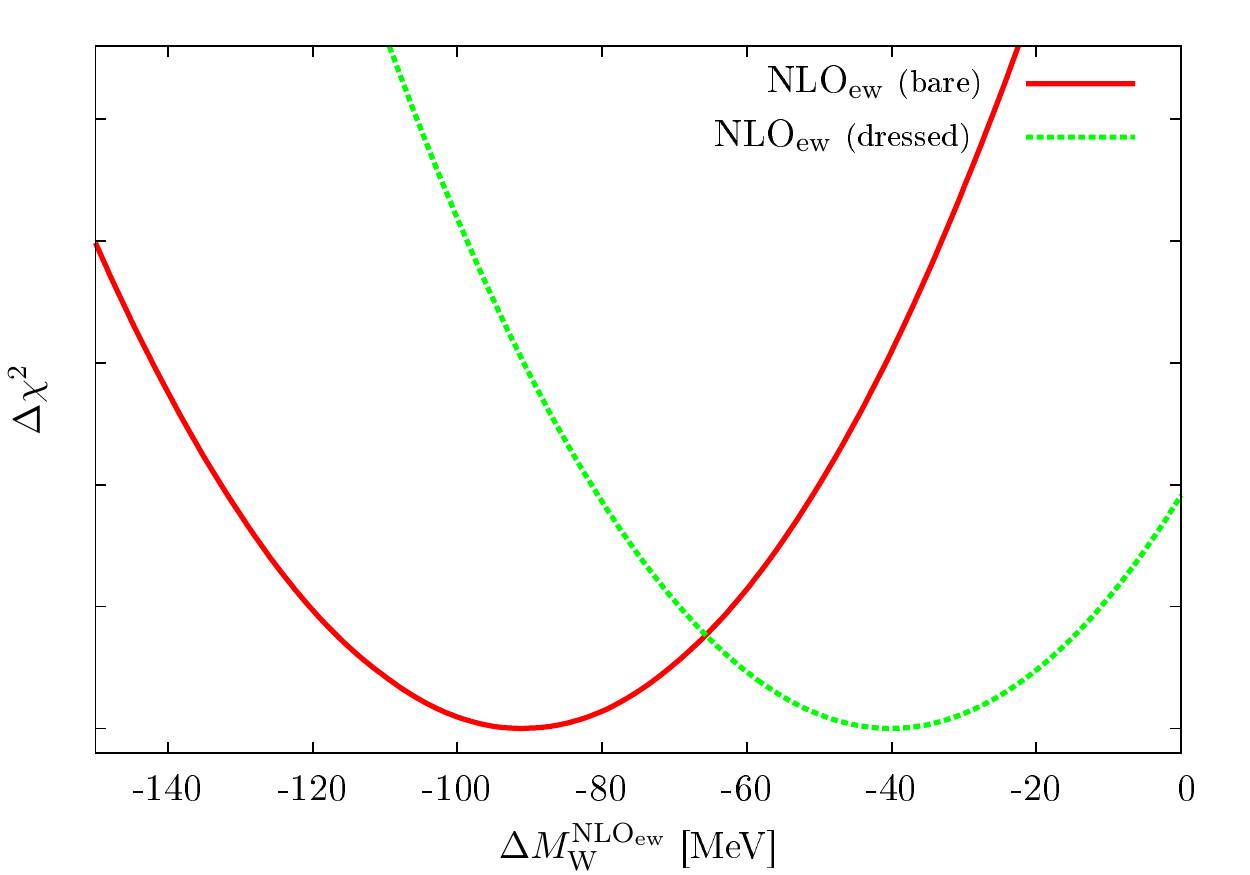}
   \includegraphics[width=.48\linewidth]{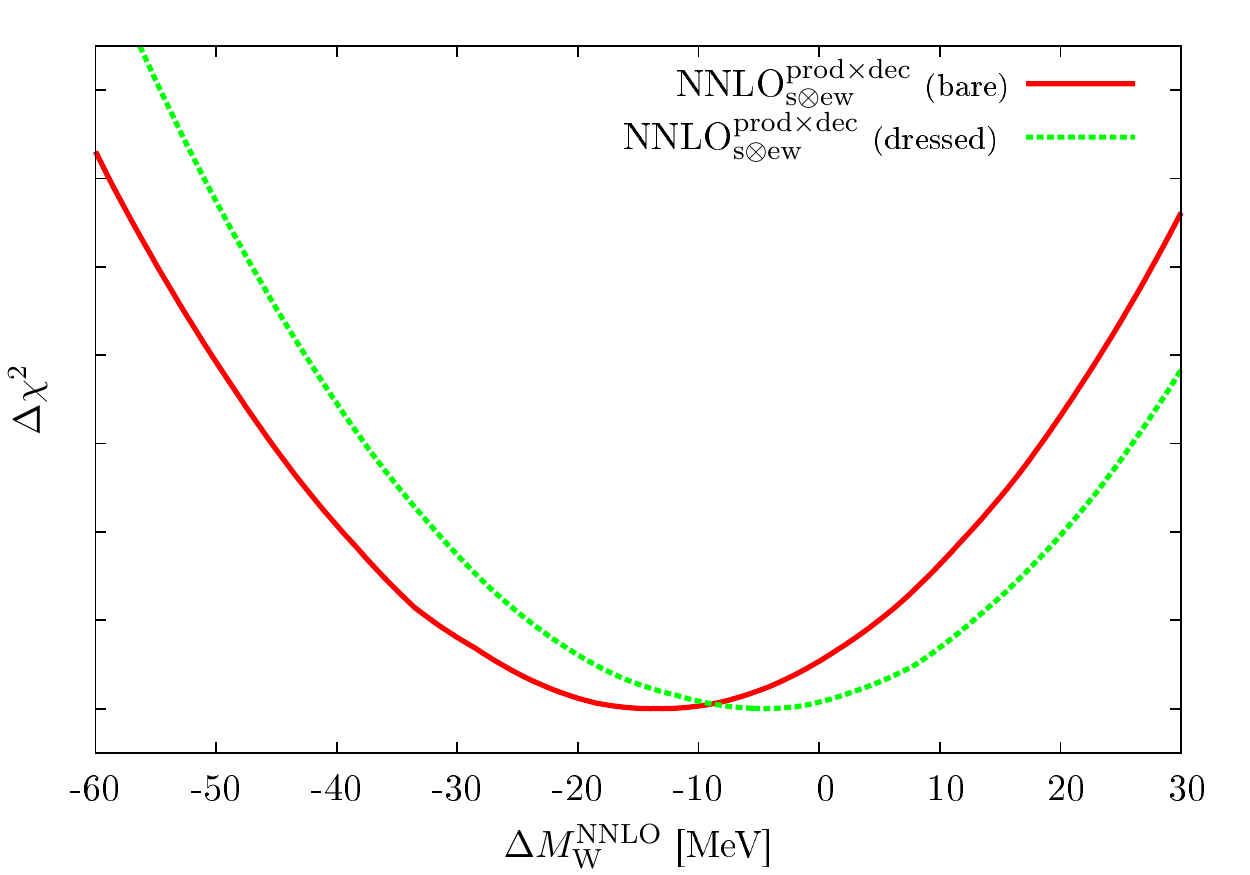}
   \caption{The $\Delta\chi^2=\chi^2-\chi^2_{\mathrm{min}}$
     distributions obtained from the fit of the NLO EW corrections
     (left) and the NNLO production--decay corrections (right) to LO
     templates in arbitrary units for $\chi^2$. The NLO mass shift $\Delta\MW^{\mathrm{NLO}_\rew}$ is given relative to the reference mass  $\MW^\OS= 80.385~\GeV$, the NNLO shift $\Delta\MW^{\mathrm{NNLO}}$ is given relative to the output mass of the fit to the sum of the EW and QCD corrections as defined in Eq~\eqref{eq:dmwnnlo}.}
   \label{fig:chi2}
 \end{figure}

In the experimental measurements of the transverse-mass distribution,
the Jacobian peak is washed out due to the finite energy and momentum
resolution of the detectors. In our simple estimate of the impact of
higher-order corrections on the extracted value of the $\PW$-boson
mass, we do not attempt to model such effects.  We expect the detector
effects to affect the different theory predictions in a similar way
and to cancel to a large extent in our estimated mass shift, which is
obtained from a difference of mass values extracted from pseudo-data
calculated using different theory predictions.  This assumption is
supported by the fact that our estimate of the effect of the NLO EW
corrections is similar to the one obtained in
Ref.~\cite{CarloniCalame:2003ux} using a Gaussian smearing of the
four-momenta to simulate detector effects.

\begin{table}
\centering
\begin{tabular}{@{} l r@{\enspace}l r@{\enspace}l @{}}
  \toprule
  & \multicolumn{2}{c}{bare muons} & \multicolumn{2}{c}{dressed leptons} \\
  \cmidrule(r){2-3} \cmidrule(l){4-5}
  & $M^\mathrm{fit}_\mathrm{W}\;[\mathrm{GeV}]$ & $\qquad\Delta M_\mathrm{W}$ 
  & $M^\mathrm{fit}_\mathrm{W}\;[\mathrm{GeV}]$ & $\qquad\Delta M_\mathrm{W}$ \\
  \midrule
  LO          & $80.385$ & \multirow{2}*{$\bigg\}\;-90~\mathrm{MeV}$} 
              & $80.385$ & \multirow{2}*{$\bigg\}\;-40~\mathrm{MeV}$} \\
         $\NLO_\rew$ & $80.295$ &                   
              & $80.345$ &                   \\
  \midrule
   $\NLO_{\rs\oplus\rew}$
        & $80.374$ & \multirow{2}*{$\biggr\}\;-14~\mathrm{MeV}$} 
              & $80.417$ & \multirow{2}*{$\biggr\}\;-4~\mathrm{MeV}$} \\
  NNLO        & $80.360$ &                   
              & $80.413$ &                   \\
  \bottomrule
\end{tabular}
\caption{Values of the $\PW$-boson Mass in  \GeV obtained from the $\chi^2$ fit
of the  $M_{\rT,\nu \ell}$ distribution in  different theoretical
approximations  to LO templates and the resulting mass shifts.}
  \label{tab:mass_shift}
\end{table}

The  fit results for several NLO approximations and our best NNLO prediction~\eqref{eq:def:nnlo:if} are given in Table~\ref{tab:mass_shift}.
To validate our procedure we estimate the mass shift due to the NLO EW
corrections by using the prediction
$\sigma^{\NLO_\rew}=\sigma^0+\Delta
\sigma^{\NLO_\rew}$
 as the pseudo-data $\sigma^{\mathrm{th}}$
in~\eqref{eq:chi2}. The $\chi^2$ distribution is shown on the left-hand side of
Fig.~\ref{fig:chi2} as a function of the mass shift
$\Delta\MW^{\NLO_\rew}$ for the dressed-lepton and bare-muon cases.
From the minima of the distributions one finds a mass shift of
$\Delta \MW^{\NLO_\rew}\approx -90~\MeV$ for bare muons and $\Delta
\MW^{\NLO_\rew}\approx -40~\MeV$ for dressed muons.  These values are
comparable to previous results reported in 
Ref.~\cite{CarloniCalame:2003ux}.\footnote{
  In Ref.~\cite{CarloniCalame:2003ux} the values $\Delta\MW=110~\MeV$
  ($20~\MeV$) are obtained for the bare-muon (dressed-lepton)
  case. These values are obtained using the
  $\order{\alpha}$-truncation of a LL shower and for
  lepton-identification criteria appropriate for the Tevatron taken
  from Ref.~\cite{Baur:1998kt}, so they cannot be compared directly to
  our results. In particular, in the dressed-lepton case, a looser
  recombination criterion $R_{\Plpm\Pgg}< 0.2$ is applied, which is
  consistent with a smaller impact of the EW corrections.  Note that
  the role of pseudo-data and templates is reversed in
  Ref.~\cite{CarloniCalame:2003ux} so that the mass shift has the
  opposite sign.}  
Alternatively, the effect of the EW corrections can
be estimated by comparing the value of $\MW$ obtained from a fit to
the naive product of EW and QCD
corrections~\eqref{eq:def:nnlo:naive:fact} to the result of a fit to
the NLO QCD cross section. The results are consistent with the shift
estimated from the NLO EW corrections alone. 

We have also estimated the effect of multi-photon radiation on the
$\MW$ measurement in the bare-muon case using the
structure-function approach given in  Eq.~\eqref{eq:LL1FSR}. As
discussed in detail in Ref.~\cite{Brensing:2007qm} we match the exponentiated LL-FSR
corrections evaluated in the  $\alpha(0)$-scheme to
the NLO calculation in the $\alpha_{G_\mu}$-scheme, avoiding double-counting. 
We obtain a mass shift $\Delta\MW^{\mathrm{FSR}}\approx 9~\MeV$
relative to the result of the fit to the NLO EW prediction, which
is in qualitative agreement with the result of
Ref.~\cite{CarloniCalame:2003ux}.

To estimate the impact of the  initial--final \order{\alphas\alpha}
corrections  we consider the mass shift relative to the full NLO result,
\begin{equation}
  \label{eq:dmwnnlo}
  \Delta\MW^{\NNLO}=\MW^{\mathrm{fit},\NNLO_{\rs\otimes\rew}^{\pro\times\dec}}
  -\MW^{\mathrm{fit},\NLO_{\rs\oplus\rew}}
\end{equation}
where $\MW^{\mathrm{fit},\NNLO_{\rs\otimes\rew}^{\pro\times\dec}}$ is the
result of using our best
prediction~\eqref{eq:def:nnlo:if} to generate the pseudo-data, while
the sum of the NLO QCD and EW corrections is used for 
$\Delta\MW^{\mathrm{fit},\NLO_{\rs\oplus\rew}}$. The resulting $\Delta\chi^2$
distributions for the mass shift are shown in the right-hand plot in
Fig.~\ref{fig:chi2}.  
In the bare-muon case, we obtain
a mass shift due to \order{\alphas\alpha} corrections of $\Delta
\MW^{\mathrm{NNLO}}\approx -14~\MeV$ while for the dressed-lepton case we
get $\Delta \MW^{\mathrm{NNLO}}\approx -4~\MeV$.  

Identical shifts result from replacing the NNLO prediction by the
naive product~\eqref{eq:def:nnlo:naive:fact}, which is expected from
the good agreement for the $M_{\rT,\nu \ell}$-spectrum in
Fig.~\ref{fig:distNNLO-IF-Wp}.  Using instead the leading-logarithmic
approximation of the final-state photon radiation obtained using
PHOTOS to compute the \order{\alphas\alpha} corrections, we obtain a
mass shift of $\Delta\MW^{\NNLO}=-11~\MeV$ ($-4~\MeV$) for the
bare-muon (dressed-lepton) case.  The effect of the
\order{\alphas\alpha} corrections on the mass measurement is therefore
of a similar or larger magnitude than the effect of multi-photon
radiation.
  We emphasize that 
the result $\Delta\MW^{\NNLO}\approx -14~\MeV$ is a simple estimate of the
 impact of the full \order{\alphas\alpha} corrections
on the $\MW$ measurement.  The order of magnitude shows that these
corrections must 
be taken into account properly in order to reach the $10~\MeV$ accuracy
goal of the LHC experiments. 
It is beyond the scope of this paper to
validate the accuracy of the 
previous and current theoretical modelling used by the
experimental collaborations in the $\MW$ measurements, which includes
the \order{\alphas\alpha} corrections in some approximation.


\section{Conclusions}
\label{sec:concl}

The Drell--Yan-like $\PW$- and $\PZ$-boson production processes are among the most precise probes of the Standard Model and do 
not only serve as key benchmark or ``standard candle'' processes, 
but further allow for precision measurements of the $\PW$-boson mass and the effective weak mixing angle.
This task of precision physics requires a further increase in the accuracy of the theoretical predictions, where the mixed QCD--electroweak corrections of \order{\alphas\alpha} currently represent the largest unknown component of radiative corrections in terms of fixed-order predictions.

In our previous paper~\cite{Dittmaier:2014qza} we have established a framework for evaluating the \order{\alphas\alpha} corrections to Drell--Yan processes in the resonance region using the pole approximation and presented the calculation of the so-called non-factorizable corrections.  They turned out to be phenomenologically negligible, so that the \order{\alphas\alpha} corrections almost entirely result from 
factorizable corrections that can be separately attributed to production and decay of the $\PW$/$\PZ$ boson (up to spin correlations).

In this paper we have presented the calculation of the 
so-called factorizable corrections of ``initial--final'' and ``final--final''
types.  The latter were calculated in Sect.~\ref{sec:calc-fact-nnlo-ff} and
only comprise finite  counterterm contributions which were found to be numerically very small ($< 0.1\%$) and therefore can be safely neglected for all phenomenological purposes.  The former, on the other hand, combine large QCD corrections to the production with large EW corrections to the decay subprocesses and are expected to be the dominant contribution of the $\order{\alphas\alpha}$ corrections.  Their calculation has been presented in Sect.~\ref{sec:calc-fact-nnlo-if}, and we have shown numerical results in Sect.~\ref{sec:if-results} for the most important observables for the $\PW$-boson mass measurement: the transverse-mass and lepton-transverse-momentum distributions for \PW production.  The results for the neutral-current process comprise the invariant-mass and the lepton-transverse-momentum distributions.  In the framework of the pole approximation, the only missing \order{\alphas\alpha} corrections are now those of ``initial--initial'' type. Based on the results of the NLO electroweak calculation, these are expected to be numerically small.

We have used our results for the dominant $\order{\alphas\alpha}$ corrections to test the validity of simpler approximate combinations of EW and QCD corrections: 
Firstly, we use a naive product ansatz multiplying the NLO QCD and EW
correction
 factors, and secondly, we approximate the $\order{\alphas\alpha}$ contribution by combining leading-logarithmic approximations of QED final-state radiation with the NLO QCD corrections.

We have demonstrated  in Sect.~\ref{sec:if-results} that naive products fail to capture the factorizable initial--final corrections in distributions such as in
the transverse momentum of the lepton, which are sensitive to QCD
initial-state radiation and therefore require  a correct treatment of the double-real-emission part of the NNLO corrections. Naive products also fail to capture observables that are strongly affected by a
 redistribution of events due to final-state real-emission corrections,
 such as the invariant-mass distribution of the neutral-current process.  On the other hand, if an observable is less affected by such a redistribution of events or is only affected by it in the vicinity of the resonance, such as the transverse-mass distribution of the charged-current process, the naive products are able to reproduce the factorizable initial--final corrections to a large extent. 

 In Sect.~\ref{sec:FSR} we have investigated to which extent the factorizable initial--final corrections calculated in this paper  can be approximated by a combination of the NLO QCD corrections and a collinear approximation of  real-photon emission through a  QED structure function approach
or a  QED parton shower such as PHOTOS.  For the invariant-mass distribution in $\PZ$-boson production we observe a significant improvement in the agreement compared to the naive product ansatz,
since
both PHOTOS and the QED structure functions model  the redistribution of events due to final-state radiation, which
 is responsible for the bulk of the corrections in this observable. 
Our results can furthermore be used to validate Monte Carlo event generators where  $\order{\alphas\alpha}$ corrections are approximated by  a combination of NLO matrix elements and parton showers. 

Finally, 
in Section~\ref{sec:mw} we have illustrated the phenomenological impact of the $\order{\alphas\alpha}$ corrections by estimating the mass shift induced by the 
 factorizable initial--final corrections as  $\approx-14~\MeV$  for the case
 of bare muons  and $\approx-4~\MeV$ 
for dressed leptons.
These corrections therefore have to be properly taken into account in the $\PW$-boson mass measurements at the LHC, which aim at a precision of about $10~\MeV$.
It will be interesting to investigate the impact of the $\order{\alphas\alpha}$ corrections on the measurement of the effective weak mixing angle as well in the future.


\section*{Acknowledgement}
This project is supported by the German Research Foundation (DFG) via grant DI 784/2-1 and the German Federal Ministry for Education and Research (BMBF). 
Moreover, A.H.\ is supported via the ERC Advanced Grant MC@NNLO (340983).
C.S.\ is supported by the Heisenberg Programme of the DFG.

\appendix
\section*{Appendix}


\section{Renormalization constants for the leptonic vector-boson decay at \texorpdfstring{$\order{\alphas\alpha}$}{O(as a)}}
\label{app:ff}

In this appendix we provide the expressions for the finite
$\mathcal{O}(\alphas\alpha)$ counterterms to the leptonic vertices of
the $\PW$ and $\PZ$ bosons in the on-shell renormalization
scheme following the conventions of Ref.~\cite{Denner:1991kt}.

\subsection{Vector-boson self-energies}

The transverse and longitudial parts of the vector-boson self-energy,
$\Sigma^{V_aV_b}_\rT$ and $\Sigma^{V_aV_b}_\rL$, are defined by the decomposition of the
irreducible two-point function $\Gamma^{V_aV_b}_{\mu\nu}$ as
\begin{equation}
  \Gamma^{V_aV_b}_{\mu\nu}(q)=-\ri \delta_{ab} g_{\mu\nu} (q^2-M_{V_a}^2)-
\ri  \left(g_{\mu\nu}-\frac{q_\mu q_\nu}{q^2}\right)\Sigma^{V_aV_b}_\rT(q^2)-\ri
  \frac{q_\mu q_\nu}{q^2}\Sigma^{V_aV_b}_\rL(q^2),
\end{equation}
where $q$ is the momentum carried by the vector bosons $V_{a,b}$.
The $\mathcal{O}(\alphas\alpha)$ corrections to the vector-boson
self-energies are given in Ref.~\cite{Djouadi:1993ss} in  terms of
scalar functions $\Pi^{V,A}_\rT$.\footnote{%
Note that in the expression given in Eq.~(5.4) of Ref.~\cite{Djouadi:1993ss} for
$\Pi^{V,A}_\rT$ in the special case of one vanishing fermion mass, the sign of the term 
$1/3(2+\alpha)(\alpha-1)G(x)$ should be reversed in agreement with Ref.~\cite{Djouadi:1987di}. We thank Paolo Gambino
for communication on this point.
} 
Treating all quarks apart from the top quark as massless, the transverse parts of the vector-boson self-energies can be expressed as follows,
\begin{align}
\Sigma_\rT^{WW, (\alphas\alpha)}(s) & =\frac{\alphas\alpha}{8\pi^2 \sw^2}
   \left[2({\Pi}^V_\rT(s,0,0)+{\Pi}^A_\rT(s,0,0))
   +   ({\Pi}^V_\rT(s,m_\Pqt^2,0)+{\Pi}^A_\rT(s,m_\Pqt^2,0))
   \right], \nonumber\\
\Sigma_\rT^{ZZ,(\alphas\alpha)}(s)&
=\frac{\alphas\alpha}{4\pi^2 \sw^2 \cw^2}
   \left[\left(\frac{44}{9}\sw^4-\frac{14}{3}\sw^2+\frac{5}{4} \right)
     {\Pi}^V_\rT(s,0,0)+\tfrac{5}{4}{\Pi}^A_\rT(s,0,0) \right. \nonumber\\
&\quad\left.   +  \left(\frac{1}{2}-\frac{4}{3} \sw^2\right)^2{\Pi}^V_\rT(s,m_\Pqt^2,m_\Pqt^2)
+\tfrac{1}{4}{\Pi}^A_\rT(s,m_\Pqt^2,m_\Pqt^2)
   \right], \nonumber \\
\Sigma_\rT^{AA,(\alphas\alpha)}(s)&
=\frac{\alphas\alpha}{\pi^2}
\left[\frac{11}{9}{\Pi}^V_\rT(s,0,0) + \frac{4}{9} {\Pi}^V_\rT(s,m_\Pqt^2,m_\Pqt^2)\right],
\nonumber \\
\Sigma_\rT^{AZ,(\alphas\alpha)}(s)
&=-\frac{\alphas\alpha}{2\pi^2\sw\cw}\left[
\left(\frac{7}{6}-\frac{22}{9}\sw^2\right)\Pi^V_\rT(s,0,0)
+\left(\frac{1}{3}-\frac{8}{9} \sw^2\right)\Pi^V_\rT(s,m_\Pqt^2,m_\Pqt^2)\right] ,
\end{align}
where the explicit expressions for the scalar functions $\Pi^{V,A}_\rT$
given in Ref.~\cite{Djouadi:1993ss} include the colour factor $N_c=3$.
The top-quark mass renormalization is performed in the on-shell
scheme.

\subsection{Vertex counterterms}
At $\order{\alphas\alpha}$, the vertex counterterms for the leptonic
vector-boson decay receive only contributions from the vector-boson
self-energies, 
\begin{align}
  \delta_{\PW\Pl_1\Pal_2}^{\mathrm{ct},(\alphas\alpha)}&=
  \delta Z_e^{(\alphas\alpha)} -\frac{\delta \sw^{(\alphas\alpha)}}{\sw}
  +\frac{1}{2}\delta Z_W^{(\alphas\alpha)},\nonumber\\
  \delta^{\mathrm{ct},\tau_\ell,(\alphas\alpha)}_{\PZ\Pl\Pal} &=
  \frac{\delta g_\Pl^{\tau_\ell,(\alphas\alpha)}}{g_\Pl^{\tau_\ell}}  
  +\frac{1}{2}\delta Z^{(\alphas\alpha)}_{ZZ} - \frac{Q_\Pl}{2g_\Pl^{\tau_\ell}}
\delta Z_{AZ}^{(\alphas\alpha)},
 \end{align}
where $\tau_\ell=\pm$ denotes the lepton chirality.
The coupling constants entering the $\PZ$-boson vertex are given by
\begin{equation}
  \begin{aligned}
    g_\Pl^+&=-\frac{\sw}{\cw}Q_\Pl, & 
    \delta g_\Pl^{+}&=-\frac{\sw}{\cw}Q_\Pl
    \left[\delta Z_e
        +\frac{1}{\cw^2}\frac{\delta \sw}{\sw}\right],  \\
    g_\Pl^-&=\frac{1}{\sw\cw}\left(I^3_{\rw,\Pl}-\sw^2 Q_\Pl\right),&
    \delta g_\Pl^{-}&=\frac{I^3_{\rw,\Pl}}{\sw\cw}
    \left[\delta Z_e
        +\frac{\sw^2-\cw^2}{\cw^2}\frac{\delta \sw}{\sw} \right]+ 
\delta g_\Pl^{+},
  \end{aligned}
\end{equation}
where $I^3_{\rw,\Pl}=-\frac{1}{2}$ is the third component of the weak
isospin of the charged lepton $\Pl$.

The vector-boson wave-function renormalization constants can be expressed in terms of the self-energies as follows,
\begin{align}
  \delta Z_{AA}&=-\left.\frac{\partial\Sigma^{AA}_\rT(k^2)}{\partial k^2}
  \right|_{k^2=0}, &&
  \nonumber\\
  \delta Z_{W}&=-\left.
\Re\frac{\partial \Sigma_\rT^{WW}(s)}{\partial s}\right|_{s=\MW^2}, & 
 \delta Z_{ZZ}&=-\left. \Re \frac{\partial \Sigma_\rT^{ZZ}(s)}{\partial s}
   \right|_{s=\MZ^2},
\nonumber \\
 \delta Z_{ZA}&=2\frac{\Sigma_\rT^{AZ}(0)}{\MZ^2}, &
 \delta Z_{AZ}&=-2\Re\frac{\Sigma_\rT^{AZ}(\MZ^2)}{\MZ^2},
\end{align}
where the $\order{\alphas\alpha}$ contribution to $ \delta Z_{ZA}$ vanishes~\cite{Djouadi:1993ss}.
The charge-renormalization constant $\delta Z_e$ in the  $\alpha(0)$ input-parameter scheme
 is given by
\begin{align}
\label{eq:dZe0}
  \delta Z_e^{\alpha(0)}
&=-\frac{1}{2}\delta Z_{AA}-\frac{\sw}{\cw}\frac{1}{2}\delta Z_{ZA}.
\end{align}
The transition to the  $G_\mu$-scheme is performed according to
Eq.~\eqref{eq:dZeGmu}. The renormalization constant for the weak mixing angle is given by
\begin{equation}
  \frac{\delta \sw}{\sw}
=-\frac{\cw^2}{2\sw^2}
\left(\frac{\Re\Sigma_\rT^{WW}(\MW^2)}{\MW^2}-\frac{\Re\Sigma_\rT^{ZZ}(\MZ^2)}{\MZ^2}\right).
\end{equation}

The final expressions of the  $\order{\alphas\alpha}$ counterterms in the $G_\mu$-scheme in terms of self-energies are then given by
\begin{align}
\delta_{\PW\Pl_1\Pal_2}^{\mathrm{ct},(\alphas\alpha)}=&
  -\frac{1}{2}\left(
    \Re\frac{\partial \Sigma_\rT^{WW,(\alphas\alpha)}(s)}{\partial s}|_{s=\MW^2} 
    +\frac{\Sigma^{WW,(\alphas\alpha)}_{\rT}(0)
      -\Re \Sigma^{WW,(\alphas\alpha)}_{\rT}(\MW^2)}{\MW^2} \right),\nonumber \\
   \delta^{\mathrm{ct},+,(\alphas\alpha)}_{\PZ\Pl\Pal} =&
-\frac{1}{2}
    \frac{\Re\partial \Sigma_\rT^{ZZ,(\alphas\alpha)}(s)}{\partial s}|_{s=\MZ^2}
    -\left(1-\frac{2}{\sw^2}\right)
    \frac{\Re\Sigma_\rT^{ZZ,(\alphas\alpha)}(\MZ^2)}{2\MZ^2}
\nonumber\\
&
-\frac{\Sigma^{WW,(\alphas\alpha)}_{\rT}(0)}{2\MW^2}
+ \left(1-\frac{1}{\sw^2}\right)
  \frac{\Re\Sigma^{WW,(\alphas\alpha)}_{\rT}(\MW^2)}{\MW^2}
  -\frac{\cw}{\sw}\frac{\Re\Sigma_\rT^{AZ,(\alphas\alpha)}(\MZ^2)}{\MZ^2},\nonumber\\
  \delta^{\mathrm{ct},-,(\alphas\alpha)}_{\PZ\Pl\Pal}=& \,
  \delta^{\mathrm{ct},+,(\alphas\alpha)}_{\PZ\Pl\Pal}
+\frac{I^3_{\rw,f}}{\sw\cw g_{\ell}^-}
\left[-\frac{2\delta \sw^{(\alphas\alpha)}}{\sw}
+\frac{\cw}{\sw}\frac{\Sigma_\rT^{AZ,(\alphas\alpha)}(\MZ^2)}{\MZ^2}\right].
\label{eq:dZVNNLO}
\end{align}


\section{Explicit form of the IR-safe contributions to the factorizable initial--final corrections}
\label{app:ir-safe}

In this section we provide the explicit expressions for each of the contributions to the factorizable initial--final corrections in our master formula~\eqref{eq:IF:final-master}.

The double-virtual corrections~\eqref{eq:IF:master-VV} and the (virtual
QCD)$\times$(real photonic) corrections~\eqref{eq:IF:master-VR}
are obtained by dressing the virtual part of the NLO QCD corrections with the factorizable final-state EW corrections,
\begin{subequations}
\label{eq:IF:master-VX-qq}
\begin{align}
  \label{eq:IF:master-VV-qq}
  \sigreg_{\Paq_a\Pq_b,\pro\times\dec}^{\Vs\otimes\Vew} 
  &=
  \int_2 \Bigl[ 2\Re\Bigl\{\deltadec\Bigr\} + \Iew \Bigr]
  \;\rd\sigma_{\Paq_a\Pq_b,\PA}^0\;
  \diptimes\Bigl[ 2 \Re\Bigl\{\delta_{\Vs}^{V\Paq_a\Pq_b}\Bigr\}+\CSI \Bigr] , 
  \\
  \label{eq:IF:master-VR-qq}
  \sigreg_{\Paq_a\Pq_b,\pro\times\dec}^{\Vs\otimes\Rew}
  &=
  \iint\limits_{2+\Pgg}
  \biggl\{ 
  \rd\sigma_\dec^\Rew 
  -4\pi\alpha \biggl[Q_{\Pl_1}(Q_{\Pl_1}-Q_{\Pl_2}) \; \dsub{\Pl_1V} \;
  \rd\sigma_{\Paq_a\Pq_b,\PA}^{0} \Bigl(\lips[\Pl_1V]{2}\Bigr)
  \nonumber\\&\qquad
  +Q_{\Pl_1}Q_{\Pl_2} \; \gsub{\Pl_1\Pal_2} \;
  \rd\sigma_{\Paq_a\Pq_b,\PA}^{0} \Bigl(\lips[\Pl_1\Pal_2]{2}\Bigr)
  +(\Pl_1\leftrightarrow\Pal_2)
  \biggr] 
  \diptimes
  \Bigl[ 2 \Re\Bigl\{\delta_{\Vs}^{V\Paq_a\Pq_b}\Bigr\}+\CSI \Bigr]
  \biggr\} ,
\end{align}
\end{subequations}
where the integrated counterpart of the QED dipoles $\Iew$ is defined in Eq.~\eqref{eq:nlo:dec:master:Iew}.
Here $\lips[IJ]{2}$ denotes the set of momenta of the two-particle phase
space after applying the momentum mapping associated to  the dipole $\gsub{IJ}$
or $\dsub{IJ}$. 
As in the NLO QCD corrections, only the quark--anti-quark induced
channel receives a non-vanishing contribution.

The IR-regularized (real QCD)$\times$(virtual EW) corrections~\eqref{eq:IF:master-RV} and the double-real corrections~\eqref{eq:IF:master-RR} are obtained by dressing the real-emission part of the NLO QCD cross sections with the final-state factorizable corrections.
Note the two-fold application of the dipole subtraction formalism in case of of the double-real corrections.
For the quark--anti-quark channel, the explicit expressions are
\begin{subequations}
\label{eq:IF:master-RX-qq}
\begin{align} 
  \label{eq:IF:master-RV-qq}
  \sigreg_{\Paq_a\Pq_b,\pro\times\dec}^{\Rs\otimes\Vew}
  &=
  \int_3 \;\Bigl[ 2\Re\Bigl\{\deltadec\Bigr\} + \Iew \Bigr]
  \nonumber\\\MoveEqLeft\times\biggl\{ \rd\sigma_{\Paq_a\Pq_b,\PA}^{\Rs}
  - \rd\sigma_{\Paq_a\Pq_b,\PA}^0 \Bigl(\lips[(\Paq_a\Pg)\Pq_b]{n}\Bigr) 
  \diptimes\CSV^{\Paq_a,\Paq_a}
  - \rd\sigma_{\Paq_a\Pq_b,\PA}^0 \Bigl(\lips[(\Pq_b\Pg)\Paq_a]{n}\Bigr) 
  \diptimes\CSV^{\Pq_b,\Pq_b}  
  \biggr\}, 
  \\
  \label{eq:IF:master-RR-qq} 
  \sigreg_{\Paq_a\Pq_b,\pro\times\dec}^{\Rs\otimes\Rew} 
  &=
  \iint\limits_{3+\Pgg}\biggl\{
  \rd\sigma_{\Paq_a\Pq_b,\pro\times\dec}^{\Rs\otimes\Rew}
  \nonumber\\&\quad
  - \rd\sigma_{\Paq_a\Pq_b,\dec}^\Rew \Bigl(\lips[(\Paq_a\Pg)\Pq_b]{2+\Pgg}\Bigr)
  \diptimes\CSV^{\Paq_a,\Paq_a}
  - \rd\sigma_{\Paq_a\Pq_b,\dec}^\Rew \Bigl(\lips[(\Pq_b\Pg)\Paq_a]{2+\Pgg}\Bigr)
  \diptimes\CSV^{\Pq_b,\Pq_b}
  \nonumber\\&\quad
  -4\pi\alpha \biggl[
  Q_{\Pl_1}(Q_{\Pl_1}-Q_{\Pl_2}) \; \dsub{\Pl_1V} \;
  \rd\sigma_{\Paq_a\Pq_b,\PA}^{\Rs} \Bigl(\lips[\Pl_1V]{3}\Bigr)
  \nonumber\\&\qquad
  +Q_{\Pl_1}Q_{\Pl_2} \; \gsub{\Pl_1\Pal_2} \;
  \rd\sigma_{\Paq_a\Pq_b,\PA}^{\Rs} \Bigl(\lips[\Pl_1\Pal_2]{3}\Bigr)
  +(\Pl_1\leftrightarrow\Pal_2)\biggr] 
  \nonumber\\&\quad
  +4\pi\alpha \biggl[
  Q_{\Pl_1}(Q_{\Pl_1}-Q_{\Pl_2}) \; \dsub{\Pl_1V} \;
  \rd\sigma_{\Paq_a\Pq_b,\PA}^{0} \Bigl(\dtilde{\Phi}^{(\Paq_a\Pg)\Pq_b}_{2,\Pl_1V}\Bigr)
  \diptimes\CSV^{\Paq_a,\Paq_a}
  \nonumber\\&\qquad
  +Q_{\Pl_1}(Q_{\Pl_1}-Q_{\Pl_2}) \; \dsub{\Pl_1V} \;
  \rd\sigma_{\Paq_a\Pq_b,\PA}^{0} \Bigl(\dtilde{\Phi}^{(\Pq_b\Pg)\Paq_a}_{2,\Pl_1V}\Bigr)
  \diptimes\CSV^{\Pq_b,\Pq_b}
  \nonumber\\&\qquad
  +Q_{\Pl_1}Q_{\Pl_2} \; \gsub{\Pl_1\Pal_2} \;
  \rd\sigma_{\Paq_a\Pq_b,\PA}^{0} \Bigl(\dtilde{\Phi}^{(\Paq_a\Pg)\Pq_b}_{2,\Pl_1\Pal_2}\Bigr)
  \diptimes\CSV^{\Paq_a,\Paq_a}
  \nonumber\\&\qquad
  +Q_{\Pl_1}Q_{\Pl_2} \; \gsub{\Pl_1\Pal_2} \;
  \rd\sigma_{\Paq_a\Pq_b,\PA}^{0} \Bigl(\dtilde{\Phi}^{(\Pq_b\Pg)\Paq_a}_{2,\Pl_1\Pal_2}\Bigr)
  \diptimes\CSV^{\Pq_b,\Pq_b}
  +(\Pl_1\leftrightarrow\Pal_2)\biggr] 
  \biggr\} .
\end{align}  
\end{subequations}
Here the phase-space kinematics obtained by the successive application
of both QCD and EW dipole mappings  is denoted by $\dtilde{\Phi}^{(ab)c}_{2,IJ}$.  
A detailed discussion of the behaviour of the individual terms of the double-real corrections in the various singular regions is given in Sect.~\ref{sec:IF:master}.
The corresponding expressions for the quark--gluon-initiated subprocesses read
\begin{subequations}
\label{eq:IF:master-RX-gq}
\begin{align}
  \label{eq:IF:master-RV-gq}
  \sigreg_{\Pg\Pq_b,\pro\times\dec}^{\Rs\otimes\Vew}
  &=
  \int_3 \Bigl[ 2\Re\Bigl\{\deltadec\Bigr\} + \Iew \Bigr]
  \biggl\{ \rd\sigma_{\Pg\Pq_b,\PA}^{\Rs}
  - \rd\sigma_{\Paq_a\Pq_b,\PA}^0 \Bigl(\lips[(\Pg\Pq_a)\Pq_b]{n}\Bigr) 
  \diptimes\CSV^{\Pg,\Paq_a}
  \biggr\}, 
  \\
  \label{eq:IF:master-RR-gq} 
  \sigreg_{\Pg\Pq_b,\pro\times\dec}^{\Rs\otimes\Rew} 
  &=
  \iint\limits_{3+\Pgg}\biggl\{
  \rd\sigma_{\Pg\Pq_b,\pro\times\dec}^{\Rs\otimes\Rew}
  - \rd\sigma_{\Paq_a\Pq_b,\dec}^\Rew \Bigl(\lips[(\Pg\Pq_a)\Pq_b]{2+\Pgg}\Bigr)
  \diptimes\CSV^{\Pg,\Paq_a}
  \nonumber\\&\quad
  -4\pi\alpha \biggl[
  Q_{\Pl_1}(Q_{\Pl_1}-Q_{\Pl_2}) \; \dsub{\Pl_1V} \;
  \rd\sigma_{\Pg\Pq_b,\PA}^{\Rs} \Bigl(\lips[\Pl_1V]{3}\Bigr)
  \nonumber\\&\qquad
  +Q_{\Pl_1}Q_{\Pl_2} \; \gsub{\Pl_1\Pal_2} \;
  \rd\sigma_{\Pg\Pq_b,\PA}^{\Rs} \Bigl(\lips[\Pl_1\Pal_2]{3}\Bigr)
  +(\Pl_1\leftrightarrow\Pal_2)\biggr] 
  \nonumber\\&\quad
  +4\pi\alpha \biggl[
  Q_{\Pl_1}(Q_{\Pl_1}-Q_{\Pl_2}) \; \dsub{\Pl_1V} \;
  \rd\sigma_{\Paq_a\Pq_b,\PA}^{0} \Bigl(\dtilde{\Phi}^{(\Pg\Pq_a)\Pq_b}_{2,\Pl_1V}\Bigr)
  \diptimes\CSV^{\Pg,\Paq_a}
  \nonumber\\&\qquad
  +Q_{\Pl_1}Q_{\Pl_2} \; \gsub{\Pl_1\Pal_2} \;
  \rd\sigma_{\Paq_a\Pq_b,\PA}^{0} \Bigl(\dtilde{\Phi}^{(\Pg\Pq_a)\Pq_b}_{2,\Pl_1\Pal_2}\Bigr)
  \diptimes\CSV^{\Pg,\Paq_a}
  +(\Pl_1\leftrightarrow\Pal_2)\biggr] 
  \biggr\} .
\end{align}
\end{subequations}
The contribution to the $\Pg\bar q_a$ channel is given in an analogous manner,
but is not spelled out explicitly.

The collinear counterterms with additional virtual EW~\eqref{eq:IF:master-CV} 
and real-photonic~\eqref{eq:IF:master-CR} corrections 
are constructed from the corresponding term of the NLO QCD corrections by dressing 
them with the respective factorizable final-state corrections,
\begin{subequations}
\label{eq:IF:master-CX-qq}
\begin{align}
  \label{eq:IF:master-CV-qq}
  \sigreg_{\Paq_a\Pq_b,\pro\times\dec}^{\Cs\otimes\Vew}
  &=
  \phantom{+}\int_0^1 \rd x\int_2
  \Bigl[ 2\Re\Bigl\{\deltadec\Bigr\} + \Iew \Bigr]
  \rd\sigma_{\Paq_a\Pq_b,\PA}^0 (xp_a,p_b)
  \diptimes(\CSK+\CSP)^{\Paq_a,\Paq_a} 
  \nonumber\\&\quad
  +\int_0^1 \rd x\int_2
  \Bigl[ 2\Re\Bigl\{\deltadec\Bigr\} + \Iew \Bigr]
  \rd\sigma_{\Paq_a\Pq_b,\PA}^0 (p_a,xp_b)
  \diptimes(\CSK+\CSP)^{\Pq_b,\Pq_b} ,
  \\
  \label{eq:IF:master-CR-qq}
  \sigreg_{\Paq_a\Pq_b,\pro\times\dec}^{\Cs\otimes \Rew}
  &=
  \int_0^1\rd x\iint\limits_{2+\Pgg}
  \biggl\{
  \rd\sigma_{\Paq_a\Pq_b,\dec}^\Rew(xp_a,p_b)
  \nonumber\\&\quad
  -4\pi\alpha \biggl[
  Q_{\Pl_1}(Q_{\Pl_1}-Q_{\Pl_2}) \; \dsub{\Pl_1V} \;
  \rd\sigma_{\Paq_a\Pq_b,\PA}^{0} \Bigl(\lips[\Pl_1V]{2}(xp_a,p_b)\Bigr)
  \nonumber\\&\quad
  +Q_{\Pl_1}Q_{\Pl_2} \; \gsub{\Pl_1\Pal_2} \;
  \rd\sigma_{\Paq_a\Pq_b,\PA}^{0} \Bigl(\lips[\Pl_1\Pal_2]{2}(xp_a,p_b)\Bigr)
  +(\Pl_1\leftrightarrow\Pal_2)\biggr] 
  \biggr\}\diptimes(\CSK+\CSP)^{\Paq_a,\Paq_a}  
  \nonumber\\&
  + \int_0^1\rd x\iint\limits_{2+\Pgg}
  \biggl\{
  \rd\sigma_{\Paq_a\Pq_b,\dec}^\Rew(p_a,xp_b)
  \nonumber\\&\quad
  -4\pi\alpha \biggl[
  Q_{\Pl_1}(Q_{\Pl_1}-Q_{\Pl_2}) \; \dsub{\Pl_1V} \;
  \rd\sigma_{\Paq_a\Pq_b,\PA}^{0} \Bigl(\lips[\Pl_1V]{2}(p_a,xp_b)\Bigr)
  \nonumber\\&\quad
  +Q_{\Pl_1}Q_{\Pl_2} \; \gsub{\Pl_1\Pal_2} \;
  \rd\sigma_{\Paq_a\Pq_b,\PA}^{0} \Bigl(\lips[\Pl_1\Pal_2]{2}(p_a,xp_b)\Bigr)
  +(\Pl_1\leftrightarrow\Pal_2)\biggr] 
  \biggr\}\diptimes(\CSK+\CSP)^{\Pq_b,\Pq_b}  
  .
\end{align}
\end{subequations}
The corresponding formulae for the gluon--quark channel read
\begin{subequations}
\label{eq:IF:master-CX-gq}
\begin{align}
  \label{eq:IF:master-CV-gq}
  \sigreg_{\Pg\Pq_b,\pro\times\dec}^{\Cs\otimes\Vew}
  &=
  \phantom{+}\int_0^1 \rd x\int_2
  \Bigl[ 2\Re\Bigl\{\deltadec\Bigr\} + \Iew \Bigr]
  \rd\sigma_{\Paq_a\Pq_b,\PA}^0 (xp_{\Pg},p_b)
  \diptimes(\CSK+\CSP)^{\Pg,\Paq_a}  ,
  \\
  \label{eq:IF:master-CR-gq}
  \sigreg_{\Pg\Pq_b,\pro\times\dec}^{\Cs\otimes \Rew}
  &=
  \int_0^1\rd x\iint\limits_{2+\Pgg}
  \biggl\{
  \rd\sigma_{\Paq_a\Pq_b,\dec}^\Rew(xp_{\Pg},p_b)
  \nonumber\\*&\quad
  -4\pi\alpha \biggl[
  Q_{\Pl_1}(Q_{\Pl_1}-Q_{\Pl_2}) \; \dsub{\Pl_1V} \;
  \rd\sigma_{\Paq_a\Pq_b,\PA}^{0} \Bigl(\lips[\Pl_1V]{2}(xp_{\Pg},p_b)\Bigr)
  \nonumber\\*&\quad
  +Q_{\Pl_1}Q_{\Pl_2} \; \gsub{\Pl_1\Pal_2} \;
  \rd\sigma_{\Paq_a\Pq_b,\PA}^{0} \Bigl(\lips[\Pl_1\Pal_2]{2}(xp_{\Pg},p_b)\Bigr)
  +(\Pl_1\leftrightarrow\Pal_2)\biggr] 
  \biggr\}\diptimes(\CSK+\CSP)^{\Pg,\Paq_a}  
\end{align}
\end{subequations}
and analogous expressions for the $\Pg\bar q_a$ channel.
Here we have made the dependence on the momenta of the incoming partons explicit in order to indicate which particle undergoes a collinear splitting with the momentum fraction given by the convolution variable $x$.

\section{Non-collinear-safe observables}
\label{app:non-collinear-safe}
In order to treat non-collinear-safe observables with respect to the final-state leptons 
$i=\Pl_1,\Pal_2$ following Ref.~\cite{Dittmaier:2008md}, 
the $n$-particle kinematics in the phase space of the subtraction function
is treated as an $(n+1)$-particle event with a collinear lepton--photon pair,
where the momentum shared between the two collinear particles is controlled
by the variable $z_{iJ}$,

\begin{subequations}
\label{eq:thetacut:kabelschacht:nnlo}
\begin{align}
  \rd\sigma^0_{\PA}\Bigl(\lips[iJ]{2}\Bigr) 
  &\;\longrightarrow\;
  \rd\sigma^0_{\PA}\Bigl(\lips[iJ]{2}\Bigr)\;
  \Theta_\cut\Bigl(\lips[iJ]{2} \;\Big\vert
  \;k_{i}=z_{iJ}\,\tilde{k}_{i}, 
  \;k=(1-z_{iJ})\,\tilde{k}_{i} \Bigr) ,
  \\
  \rd\sigma^\Rs_{\PA}\Bigl(\lips[iJ]{3}\Bigr) 
  &\;\longrightarrow\;
  \rd\sigma^\Rs_{\PA}\Bigl(\lips[iJ]{3}\Bigr)\;
  \Theta_\cut\Bigl(\lips[iJ]{3} \;\Big\vert
  \;k_{i}=z_{iJ}\,\tilde{k}_{i}, 
  \;k=(1-z_{iJ})\,\tilde{k}_{i} \Bigr) ,
  \\
  \rd\sigma^0_{\PA}\Bigl(\dtilde{\Phi}^{(ab)c}_{2,iJ}\Bigr)
  &\;\longrightarrow\;
  \rd\sigma^0_{\PA}\Bigl(\dtilde{\Phi}^{(ab)c}_{2,iJ}\Bigr)\;
  \Theta_\cut\Bigl(\dtilde{\Phi}^{(ab)c}_{2,iJ} \;\Big\vert
  \;\tilde{k}_{i}=z_{iJ}\,\dtilde{k}_{i}, 
  \;\tilde{k}=(1-z_{iJ})\,\dtilde{k}_{i} \Bigr) ,
\end{align}
\end{subequations}
where we have made explicit the cut function for the computation of observables in the notation.
This modification induces additional convolution terms over the distribution $[\cIbarew(z)]_+$ with
\begin{align}
  \label{eq:nlo:dec:master:cIbarew}
  \cIbarew(z)
  &=
  \frac{\alpha}{2\pi}\, Q_{\Pl_1}\biggl[
  (Q_{\Pl_1}-Q_{\Pl_2})\; \cDbarsub{\Pl_1 V}(z)
  +Q_{\Pl_2}\; \cGbarsub{\Pl_1\Pal_2}(z)
  \biggr]
  \nonumber\\&\quad\times
  \Theta_\cut\left(\lips[]{2} \;\Big\vert
  \;k_{\Pl_1}=z\,\tilde{k}_{\Pl_1}, 
  \;k=(1-z)\,\tilde{k}_{\Pl_1} \right)
  +(\Pl_1\leftrightarrow\Pal_2) ,
\end{align}
which we indicate by the label ``$\Rbarew$''.
The contribution with virtual QCD corrections is given by
\begin{align}
  \label{eq:IF:master-VRbar-qq}
  \sigreg_{\Paq_a\Pq_b,\pro\times\dec}^{\Vs\otimes\Rbarew} 
  &=
  \int_0^1\rd z \int_2 \left[\cIbarew(z)\right]_+ 
  \;\rd\sigma_{\Paq_a\Pq_b,\PA}^0\;
  \diptimes\Bigl[ 2 \Re\Bigl\{\delta_{\Vs}^{V\Paq_a\Pq_b}\Bigr\}+\CSI \Bigr] , 
\end{align}
and the real-emission corrections for the quark--anti-quark and gluon--quark induced contributions read
\begin{align} 
  \label{eq:IF:master-RRbar-qq}
  \sigreg_{\Paq_a\Pq_b,\pro\times\dec}^{\Rs\otimes\Rbarew}
  &=
  \int_0^1\rd z \int_3 \left[\cIbarew(z)\right]_+ 
  \biggl\{ \rd\sigma_{\Paq_a\Pq_b,\PA}^{\Rs}
  - \rd\sigma_{\Paq_a\Pq_b,\PA}^0 \Bigl(\lips[(\Paq_a\Pg)\Pq_b]{n}\Bigr) 
  \diptimes\CSV^{\Paq_a,\Paq_a}
  \nonumber\\&\quad
  - \rd\sigma_{\Paq_a\Pq_b,\PA}^0 \Bigl(\lips[(\Pq_b\Pg)\Paq_a]{n}\Bigr) 
  \diptimes\CSV^{\Pq_b,\Pq_b}  
  \biggr\} ,
\end{align}
\begin{align}
  \label{eq:IF:master-RRbar-gq}
  \sigreg_{\Pg\Pq_b,\pro\times\dec}^{\Rs\otimes\Rbarew}
  &=
  \int_0^1\rd z \int_3 \left[\cIbarew(z)\right]_+ 
  \biggl\{ \rd\sigma_{\Pg\Pq_b,\PA}^{\Rs}
  - \rd\sigma_{\Paq_a\Pq_b,\PA}^0 \Bigl(\lips[(\Pg\Pq_a)\Pq_b]{n}\Bigr) 
  \diptimes\CSV^{\Pg,\Paq_a}
  \biggr\} 
\end{align}
and analogous terms for the $\Pg\bar q_a$ channel.
The $\CSK$ and $\CSP$ operators contain an additional convolution over the momentum 
fractions~$x$ of the incoming partons and can be written in the following form,
\begin{align}
  \label{eq:IF:master-CRbar-qq}
  \sigreg_{\Paq_a\Pq_b,\pro\times\dec}^{\Cs\otimes\Rbarew}
  &=
  \phantom{+}\int_0^1 \rd x \int_0^1\rd z  \int_2  \left[\cIbarew(z)\right]_+
  \rd\sigma_{\Paq_a\Pq_b,\PA}^0 (xp_a,p_b)
  \diptimes(\CSK+\CSP)^{\Paq_a,\Paq_a} 
  \nonumber\\&\quad
  +\int_0^1 \rd x \int_0^1\rd z  \int_2 \left[\cIbarew(z)\right]_+
  \rd\sigma_{\Paq_a\Pq_b,\PA}^0 (p_a,xp_b)
  \diptimes(\CSK+\CSP)^{\Pq_b,\Pq_b} ,
\end{align}
\begin{align}
  \label{eq:IF:master-CRbar-gq}
  \sigreg_{\Pg\Pq_b,\pro\times\dec}^{\Cs\otimes\Rbarew}
  &=
  \phantom{+}\int_0^1 \rd x \int_0^1\rd z  \int_2 \left[\cIbarew(z)\right]_+ 
  \rd\sigma_{\Paq_a\Pq_b,\PA}^0 (xp_{\Pg},p_b)
  \diptimes(\CSK+\CSP)^{\Pg,\Paq_a} 
\end{align}
and analogous terms for the $\Pg\bar q_a$ channel.
Note that the $\CSK$ and $\CSP$ operators, in general, contain plus distributions with respect to the variable $x$, and the above equations need to be properly evaluated in combination with the plus distribution $[\cIbarew(z)]_+$ that acts on the integration variable~$z$.


\section{PHOTOS settings}
\label{app:photos}

The results using the PHOTOS parton shower shown in Figs.~\ref{fig:distNNLO-IF-FSR-Wp} and \ref{fig:distNNLO-IF-FSR-Z} of Sect.~\ref{sec:FSR} were obtained with version 2.15 of the program and the following options:
\begin{tabbing}
  \qquad\=\hspace{.35\textwidth}\=\hspace{.35\textwidth}\=\kill
  \> \texttt{ISEC=.FALSE.}, 
  \> \texttt{ITRE=.FALSE.}, 
  \> \texttt{IEXP=.FALSE.}, \\
  \> \texttt{IFTOP=.FALSE.}, 
  \> \texttt{XPHCUT=0.01D0} \; (default value).
\end{tabbing}
These settings restrict the parton shower to \emph{at most one} additional photon emission in order to simulate the impact of $\order{\alpha}$ corrections. 
Further settings which differ for the charged-current and neutral-current processes are as follows:
\begin{tabbing}
  \qquad\=\hspace{.35\textwidth}\=\hspace{.35\textwidth}\=\kill
  \> \textbf{\boldmath{\large \PWpm production:}} \> \textbf{\boldmath{\large \PZ production:}} \\
  \> \texttt{IFW=.TRUE.},              \> \texttt{IFW=.FALSE.}, \\
  \> \texttt{INTERF=.FALSE.},          \> \texttt{INTERF=.TRUE.}, \\
  \> \texttt{ALPHA=}\;$\alpha_{G_\mu}$, \> \texttt{ALPHA=}\;$\alpha(0)$.
\end{tabbing}
Note that the electromagnetic coupling constant $\alpha$ is adjusted to the respective value used in our calculation of the initial--final corrections as described in Sect.~\ref{sec:IF:numerics}.

\bibliographystyle{tep}
\bibliography{dyfact}

\end{document}